\documentclass[iop,numberedappendix,twocolappendix]{emulateapj}

\notetoeditor{Please keep the plots rotated/trimmed/clipped as they are here}

\DeclareRobustCommand{\rchi}{{\mathpalette\irchi\relax}}
\newcommand{\irchi}[2]{\raisebox{\depth}{$#1\chi$}} 

\usepackage{tabularx}
\usepackage{threeparttable}
\usepackage{multirow}
\renewcommand{\arraystretch}{1.45}

\usepackage{amssymb}
\usepackage{amsmath}

\usepackage{graphicx,subfigure}

\usepackage{url}
\usepackage{verbatim}

\usepackage{natbib}
\usepackage{xcolor}
\definecolor{grassy}{rgb}{0.1,0.7,0.2}
\usepackage[colorlinks=true]{hyperref}
\hypersetup{
    colorlinks,	
    citecolor=grassy,
}

\newcommand{\noteR}[1]{{\color{red}\bf{#1}}}					

\shorttitle{Testing the Kerr metric with Mrk~335 X-ray Reflection}
\shortauthors{Choudhury et al.}

\begin{document}

\title{Testing the Kerr metric with X-ray Reflection Spectroscopy \\ of Mrk~335 \textit{Suzaku} data}


\author{Kishalay~Choudhury\altaffilmark{1}}
\author{Sourabh~Nampalliwar\altaffilmark{2}}
\author{Askar~B.~Abdikamalov\altaffilmark{1}}
\author{Dimitry~Ayzenberg\altaffilmark{1}}
\author{Cosimo~Bambi\altaffilmark{1,2,$*$,$\dagger$}}
\author{Thomas~Dauser\altaffilmark{3}}
\author{Javier~A.~Garc{\'{\i}}a\altaffilmark{4,3,$\dagger$}}

\altaffiltext{1}{Center for Field Theory and Particle Physics and Department of Physics, Fudan University, 200438 Shanghai, China}

\altaffiltext{2}{Theoretical Astrophysics, Eberhard-Karls Universit{\"a}t T{\"u}bingen, 72076 T{\"u}bingen, Germany}

\altaffiltext{3}{Remeis Observatory \& ECAP, Universit{\"a}t Erlangen-N{\"u}rnberg, 96049 Bamberg, Germany}

\altaffiltext{4}{Cahill Center for Astronomy and Astrophysics, California Institute of Technology, Pasadena, CA 91125, USA}

\altaffiltext{$*$}{\email[Corresponding~author: ]{bambi@fudan.edu.cn}}
\altaffiltext{$\dagger$}{Alexander~von~Humboldt Fellow}

\begin{abstract}

Einstein's gravity has undergone extensive tests in the weak field gravitational limit, with results in agreement with theoretical predictions. There exist theories beyond general relativity (GR) which modify gravity in the strong field regime but agree with GR in the weak field. Astrophysical black holes are believed to be described by the Kerr metric and serve as suitable candidates to test strong gravity with electromagnetic radiation. We perform such a test by fitting one \textit{Suzaku} dataset of the narrow-line Seyfert 1 (NLS1) galaxy Mrk~335 with X-ray reflection spectroscopy, using the Johannsen metric to model the black hole spacetime and test for deviations from Kerr. We find the data is best modeled with a hybrid model that includes both partial covering absorption and a reflection component. This is the first time such a model has been proposed for a high-flux (low reflection) Mrk~335 dataset. We constrain the Johannsen deformation parameter $\alpha_{13}$ to $-1.5<\alpha_{13}<0.6$ with spin parameter $a_{*}>0.8$, and the $\alpha_{22}$ parameter to $-0.4<\alpha_{22}<2.1$ with $a_{*}>0.7$, both at the $99\%$ confidence level. Although additional solutions at large deviations from the Kerr metric show statistical similarity with the ones above, further analysis suggests these solutions may be manifestations of uncertainties beyond our control and do not represent the data. Hence, our results are in agreement with the idea that the supermassive compact object at the center of Mrk~335 is described by the Kerr metric.

\end{abstract}

\keywords{black hole physics -- galaxies: active -- methods: data analysis -- stars: black holes -- techniques: imaging spectroscopy -- X-rays: individual (Mrk~335)}

\maketitle

\renewcommand{\thefootnote}{\roman{footnote}}

\section{Introduction} \label{intro}

	\setcounter{footnote}{0} 
	
Over a hundred years old, Albert Einstein's theory of general relativity has been tested with observations since it was proposed and agreement has been found in numerous cases spanning across a multitude of domains, notably in weak field experiments of the Solar System and for radio observations of binary pulsars \citep{Will2014}. Testing gravity in the strong field regime has gained popularity among both the electromagnetic (EM) radiation and the gravitational wave community, with astrophysical black holes proving to be the perfect candidates for carrying out such tests \citep{Yunes2013,Yagi2016,Johannsen2016,Bambi2016b,Bambi2017a,Bambi2017b}. 

In four-dimensional Einstein gravity, the Kerr metric is the only vacuum black hole solution of the field equations which is regular on and outside the event horizon, under standard assumptions like stationarity and asymptotic flatness; a consequence of the no-hair theorem \citep{Kerr1963,Carter1971,Robinson1975}. General consensus is that the Kerr metric describes the spacetime around astrophysical black holes \citep{Price1972,Bambi2009,Bambi2014}. But a number of alternative theories predict macroscopic deviations from the Kerr spacetime \citep{Mathur2005,Dvali2011,Giddings2014,Berti2015,Giddings2017}. This makes it imperative to conduct tests on the Kerr hypothesis. Observations of the X-ray reflection spectrum from the black hole neighborhood are particularly interesting as they can be used to test strong gravity by studying radiation emitted from regions very close to the black hole \citep{Fabian2000}. This method has been developed over the years assuming the Kerr metric describes the spacetime around the central compact object, in both active galactic nuclei (AGN) and black hole binaries (BHB). Blurred and distorted emission features can be seen around the reflecting regions of the accretion disk due to relativistic effects, leading to measurements of inner disk radius and black hole spin \citep{Fabian1989,Laor1991,BR2006,Rey2008}. The most notable and interesting feature is the Fe-\textrm{K$\alpha$} emission complex \citep[e.g.,][]{LW1988,GR1988,Fabian1989,GF1991}. The utilization of X-ray reflection spectroscopy for strong gravity tests has been examined in the last decade \citep{Schee2009,JP2013,Bambi2013,Jiang2015a,Jiang2015b,Bambi2016a,
Zhou2016,Ni2016,Nampalliwar2018}.

The X-ray blurring code \textsc{relxill} is currently the most popular relativistic reflection model in use that describes the reflection spectrum of optically thick, geometrically thin accretion disks \citep{ThinDiskNT1973} around black holes \citep{Garcia2014}. The code is the combination of the convoluted (\textsc{relconv}) flavor of the \textsc{relline} relativistic smearing model \citep{Dauser2010,Dauser2013} and the emission angle-dependent, non-relativistic, local disk reflection code \textsc{xillver} \citep{GK2010,Garcia2013}. \textsc{relconv} was recently modified by some of us to allow for the use of a non-Kerr metric for the purpose of testing gravity \citep{Bambi2017b}. Our relativistic blurring code \textsc{relxill\_nk} has been designed to incorporate any well-behaved, stationary, axisymmetric, and asymptotically-flat spacetimes, including parametrized metrics that deform Kerr and solutions in modified gravity theories. Recent results obtained with the code on quantifying possible Kerr deviations with X-ray data of multiple sources have been summarized in \cite{Bambi2018}. The first non-Kerr metric implemented in \textsc{relxill\_nk} was the axisymmetric metric proposed in~\cite{Johannsen2013}. Additional metrics implemented in the framework include the conformal metric of~\cite{Bambi2016c} (See \citealt{Zhou2018}) and the axisymmetric metric of Konoplya, Rezzolla \& Zhidenko in \cite{KRZ2016} (See \citealt{NampalliwarKRZ2019}). 


The paper is structured as follows-- we review the Johannsen metric (expressed in the convention assuming the natural units $G_{\textrm{N}}=c=1$) in \S\ref{metric} for the relativistic blurring code used. Our source is presented in \S\ref{source}, with the choice of the dataset described in \S\ref{Obsr.}. We explain the data reduction methodology in \S\ref{red}, followed by listing and briefly explaining the model components used for the data analysis in \S\ref{mod}. Justification of components and key observations leading to our presented results in \S\ref{sec:results} are discussed in \S\ref{discuss}. Finally, we express our concluding remarks and mention possible shortcomings in \S\ref{conc}.

\section{\textsc{relxill\_nk}: The Metric} \label{metric}

One parametrically-deformed metric to test the Kerr hypothesis is the Johannsen metric \citep{Johannsen2013} that we implement in our code \textsc{relxill\_nk}. Note that the metric is neither a solution of Einstein's field equations nor of any well-motivated modified theory of gravity. We can, however, consider it as a phenomenological hypothesis and conduct strong field tests of the no-hair theorem in general classes of gravity theories.

The line element of the Johannsen metric in Boyer-Lindquist coordinates, with the convention ($- + + +$), is given as:
\begin{widetext}
\begin{align}
ds^2       =       &-\frac{\tilde{\Sigma}\left(\Delta - a^2 A_2^{2}~\sin^2\theta\right)}{\left[\left(r^2+a^2\right)A_1 - a^2A_2\sin^2\theta\right]^2}~dt^2 -\frac{2a\left[\left(r^2+a^2\right)A_1A_2 - \Delta\right]\tilde{\Sigma}\sin^2\theta}{\left[\left(r^2+a^2\right)A_1 - a^2A_2\sin^2\theta\right]^2}~dtd\phi     \nonumber \\
                        &+ \frac{\tilde{\Sigma}}{\Delta A_5}~dr^2 + \tilde{\Sigma} ~d\theta^2    + \frac{\left[\left(r^2+a^2\right)^2~A_1^{2} - a^2\Delta\sin^2\theta\right]\tilde{\Sigma}\sin^2\theta}{\left[\left(r^2+a^2\right)A_1 - a^2A_2\sin^2\theta\right]^2}~d\phi^2  
 \label{JM}
\end{align}
\end{widetext}
where,
\begin{align}                       
A_1(r) = & 1 + \sum_{n=3}^{\infty} \alpha_{1n}\left(\frac{M}{r}\right)^n  \nonumber \\ 
A_2(r) = & 1+ \sum_{n=2}^{\infty} \alpha_{2n}\left(\frac{M}{r}\right)^n   \nonumber \\ 
A_5(r) = & 1+ \sum_{n=2}^{\infty} \alpha_{5n}\left(\frac{M}{r}\right)^n  \nonumber \\
\Delta \equiv & r^2-2Mr+a^2  \nonumber \\
\tilde{\Sigma} = & r^2+a^2\cos^2\theta + f(r) \hspace{2mm}  \nonumber \\
 f(r) = & \sum_{n=3}^{\infty}\epsilon_n\left(\frac{M^n}{r^{n-2}}\right) 
\label{DEVS}
\end{align}
with the mass $M$ and the spin parameter $a =J/M$ of the black hole, where $J$ is the spin angular momentum of the black hole. When the deviation functions $A_1=A_2=A_5=1$ and $f(r)=0$, Eq.~\ref{JM} reduces to the Kerr metric. We test for one deformation parameter at a time, keeping all others identically zero. Thus, we present two sets of results, one each for $\alpha_{13}$ and $\alpha_{22}$. We pick $\alpha_{13}$ and $\alpha_{22}$ since they are expected to have the strongest impact on the reflection spectrum \citep{Bambi2017b}, thus the constraints will be the strongest on them.

Note that there exists a degeneracy between deformation parameters (and among deformation parameters and other model parameters) in terms of their effect on the relativistic blurring, and a systematic study of various kinds of degeneracies in currently underway. But allowing for multiple simultaneous deformation parameters in our reflection model is currently computationally unfeasible owing to memory issues with the analysis package and huge FITS files requirements.

In order to avoid pathologies in the spacetime, the following limits on $\alpha_{13}$ and $\alpha_{22}$ are imposed \citep{Johannsen2013,Bambi2017b}:
\begin{align}
   \hspace{7mm} \alpha_{13} ~> -\frac{1}{2}&\left(1+\sqrt{1 - a_*^2}\right)^4 \label{13low} \\ 
    -\left(1+\sqrt{1 - a_*^2}\right)^2 < ~& \alpha_{22} ~< \frac{\left(1+\sqrt{1 - a_*^2}\right)^4}{a_*^2} \label{22} 
\end{align}
where, $a_* = a/M$ is the dimensionless spin parameter present in our code.


\section{Data Analysis} \label{analysis}

\subsection{The Source} \label{source}

The narrow-line Seyfert 1 (NLS1) AGN Mrk~335 ($z=0.0258$) has been found to host a supermassive black hole at its center with reverberation-mapped mass M$_{\bullet} \approx 2.6\times10^7$~M$_{\odot}$ \citep{Grier2012}. First detected in X-rays by UHURU \citep{Tanambaum1978}, Mrk~335 has been observed and studied numerous times by various X-ray observatories like \textit{ASCA}, \textit{Swift}, \textit{Suzaku}, \textit{XMM Newton} and \textit{NuSTAR} \citep[e.g.,][]{Ballantyne2001,Gondoin2002,Crummy2006,Grupe2007,Longinotti2007b,
ONeill2007,Larsson2008,Grupe2008,Patrick2011,Grupe2012,Gallo2013,W13,Longinotti2013,Parker2014,
Gallo2015,WG2015,Keek2016,Beheshtipour2017,Ballantyne2017,Gallo2019}. It is an extremely variable source, exhibiting more than a factor of 10 fluctuation in the X-ray flux over the past 15~years. The source has been confirmed to have a Compton reflection component, and a strong soft excess and Fe-\textrm{K$\alpha$} line broadening. Distinguishing Mrk~335 data at low energies ($<10$~keV) between partial absorption and relativistic reflection was a challenge until the first pressing evidence for relativistic reflection from the accretion disk in the AGN was presented by \cite{Kara2013}, by studying X-ray reverberation time lags. AGN exhibit highest complexity in low-flux states.

\subsection{Data} \label{Obsr.} 

Tab.~\ref{T-obs} presents a list of Mrk~335 data available online for \textit{Suzaku}, \textit{XMM Newton} and \textit{NuSTAR} missions. Mrk~335 was observed in a high-flux state on June 21, 2006 (Obs. ID: 701031010) with \textit{Suzaku} \citep{Mitsuda2007} for an X-ray Imaging Spectrometer (XIS, \citealt{Koyama2007}) net exposure time of 151~ks. It is this dataset that we will analyze in the present work owing to its least complexity, good photon counting statistics and the observation of relativistic reflection in it. The chosen observation uses \cite{W13} as a basis, which is a survey paper investigating relativistic disk reflection for 25 ``bare'' type 1 AGN \textit{Suzaku} data with little or no complicated intrinsic absorption. The cited work presented findings using the chosen Mrk~335 dataset with blurred reflection and showed the presence of relativistically-skewed \textrm{Fe} line and a Compton turnover at E~$>10$~keV. 

We compare other data in Tab.~\ref{T-obs} against our choice here. Each individual front-illuminated (FI) XIS spectrum (\S\ref{red}) in our high-flux observation has an effective exposure time ($\sim$133~ks) similar to the total exposure of the \textit{XMM Newton} high-flux 2006 dataset instead, giving us better counts with \textit{Suzaku}. An added detector advantage of \textit{Suzaku} XIS is that it had relatively lower intrinsic background contamination compared to EPIC cameras \citep{Jansen2001}, by construction.  \cite{Gallo2013}, with a combined $\sim$200~ks 2009 \textit{XMM Newton} exposure time, show weak constraints on $a_*$ (similar to \citealt{W13}) with a deeper analysis of the intermediate-flux data. The remaining datasets (all in low-flux state) have much lower effective exposures and net counts for both satellite missions, in addition to significant warm absorber modification (likely to heighten inter-component degeneracies) of the reflection data. Flaring in the $\sim$140~ks 2015 \textit{XMM Newton} low-flux observations results in a $\sim$20~ks cut-down, and the remaining observation is photon-starved \citep{Gallo2019}.

\begingroup
\begin{table}[!t]
\def\arraystretch{1.3}
\setlength{\tabcolsep}{1mm}
     \begin{ruledtabular}
     \caption{Mrk~335 archived X-ray data studied in literature for select missions.}
     
     \begin{tabular}{cccc} \label{T-obs}
     
     \multirow{2}*{Instrument} & \multirow{2}*{Obs. ID} &  \multirow{2}*{Date} & \multirow{2}*{Exposure Time} \\ [1mm]  
 & & [yyyy/mm/dd] & [ksec] \\ [1mm]\hline \hline
      \multirow{3}*{\textit{Suzaku}} & 701031010 & 2006/06/21 & 151.3 \\
      												& 708016010 & 2013/06/11 & 144.0 \\
 													& 708016020 & 2013/06/14 & 154.8 \\ \hline 
											
     \multirow{7}*{\textit{XMM Newton}} & 0101040101 & 2000/12/25 & 36.9  \\
     														& 0101040701 & 2000/12/25 & 10.9 \\
     														& 0306870101 & 2006/01/03 & 133.2 \\
 															& 0510010701 & 2007/07/10 & 22.6 \\
 															& 0600540601 & 2009/06/11 & 132.3 \\
 															& 0600540501 & 2009/06/13 & 82.6 \\
 															& 0741280201 & 2015/12/30 & 140.4 \\ \hline  
	
     \multirow{3}*{\textit{NuSTAR}} & 60001041002 & 2013/06/13 & 21.3/21.5 \\
     \multirow{3}*{(FPMA/B)}			& 60001041003 & 2013/06/13 & 21.5/21.3 \\
 													& 60001041005 & 2013/06/25 & 93.0/92.9 \\
 													& 80001020002 & 2014/09/20 & 68.9/68.8 \\ 
  											
       \end{tabular}
       \end{ruledtabular}
\end{table}
\endgroup

It is essential to state that there exist simultaneous high-energy exposures of Mrk~335 with \textit{NuSTAR} \citep{nustar2013}, which encompasses the full reflection band when used with low-energy satellite data. The only simultaneous observations available as of April 2019 (Tab.~\ref{T-obs}) are with the low-flux 2013 \textsc{Suzaku} exposures. Although these \textit{Suzaku} observations amount to $\sim300$~ks net exposure over $\sim7.7$~days, there is a threefold issue that builds up problems in using these observations: 1) The observations are photon-starved with $< 8\times 10^4$~FI XIS counts \citep{Gallo2015}; 2) The \textit{NuSTAR} coverage is short, corresponds to a very small part of the wide \textit{Suzaku} exposures, and has incredibly low statistics \citep{Parker2014}; 3) A $\sim90-100$~ks flare $\sim250$~ks into the \textit{Suzaku} observations softens the source spectrum \citep{WG2015}, thereby further reducing good intervals for a simultaneous low-flux study, especially creating problems with time-averaged analysis. A simultaneous analysis of Mrk~335 with \textit{NuSTAR}, to date, is not feasible to study strong gravity with reflection spectroscopy owing to lack of decent coverage. 

The struggle with poor detection at the low fluxes with this highly variable source is likely to make it difficult to constrain sensitive parameters such as the deformation parameters. The aim of this work is to study the constraints on possible deviations away from the Kerr metric using data quality with least complexity to avoid complicated inter-component degeneracies that may shadow our findings. In general, inclusion of other observations may provide more data. But this does not necessarily imply better constraints, since at some point systematic uncertainties dominate over statistical uncertainties (See e.g.,~\citealt{Xu2018,TripathiMCG2019tbp,Zhang2019}).


\subsection{Reduction} \label{red}

\textsc{heasoft}~v6.25 reduction and analysis package was used to process unfiltered event files of the XIS CCDs following the \textit{Suzaku} Data Reduction Guide\footnote{\url{https://heasarc.gsfc.nasa.gov/docs/suzaku/analysis/abc/}}, using XIS CALDB \texttt{v20160607}. Task \texttt{aepipeline} was run to create cleaned event files for the FI CCDs XIS0, XIS2 (non-operational since November 2006), and XIS3. Back-illuminated (BI) CCD XIS1 was not used since its sensitivity is relatively low at Fe K energies. Source (on-center) and background regions of 3.5~arcmin radii were extracted for each FI CCD on \texttt{SAOImage DS9} imaging and data visualization application.\footnote{\url{http://ds9.si.edu/site/Home.html}} Backgrounds were selected from the same CCD by avoiding the source and the $^{55}$Fe calibration sources at the corners of the CCD. Unbinned source and background spectra for each CCD were extracted using the \textsc{xselect} tool, ensuring the cutoff-rigidity was set $>6$~GeV \citep{W13} to account for proper non X-ray background (NXB) subtraction. The redistribution matrix file (RMF) and the ancillary response file (ARF) were created using the tools \textsc{xisrmfgen} and \textsc{xisarfgen},\footnote{\url{ftp://legacy.gsfc.nasa.gov/suzaku/doc/xrt/suzakumemo-2011-01.pdf}} respectively. Since we are interested in average spectral properties, the FI CCD spectra and responses were co-added using the \textsc{ftool addascaspec}. The spectra and response files were physically rebinned using the tools \textsc{rbnpha} and \textsc{rbnrmf}, respectively, with the variable binning scheme shown in the aforementioned guide. On top of this, to reduce bias imposed by a minimum grouping technique \citep[e.g., see Fig.~7 in][]{K17} and ensure high signal-to-noise ratio (SNR), we grouped our time-averaged XIS dataset to 100~cts~bin$^{-1}$ using the \textsc{ftool grppha}. Apart from data below 0.6~keV and above 10~keV, that between 1.7--2.5~keV were also ignored due to uncertainties in detector calibration around Si K edge. The resulting dataset had $\approx 5\times10^5$ total photon counts between 0.6--10~keV with a net count rate of $1.222 \pm 0.002$~s$^{-1}$ and a very low background contamination (1.2\%).

HXD/PIN \citep{Takahashi2007} data was also reduced similarly: employing \texttt{aepipeline} and then the \textsc{ftool hxdpinxbpi} using latest CALDB \texttt{v20110913}. Evidence of a Compton hump was seen, with a turnover $\sim$20~keV. However, owing to poor statistics ($\sim$6\% of total counts between 0.6--25~keV in PIN) and high errors on the data, in addition to no significant contribution to the analysis of the reflection spectrum, the PIN dataset was not included for our test of gravity.

\subsection{Modeling} \label{mod}
\begin{figure}[!t] 
\centering
\includegraphics[width=0.475\textwidth ,trim={3mm 3mm 0 0 },clip]{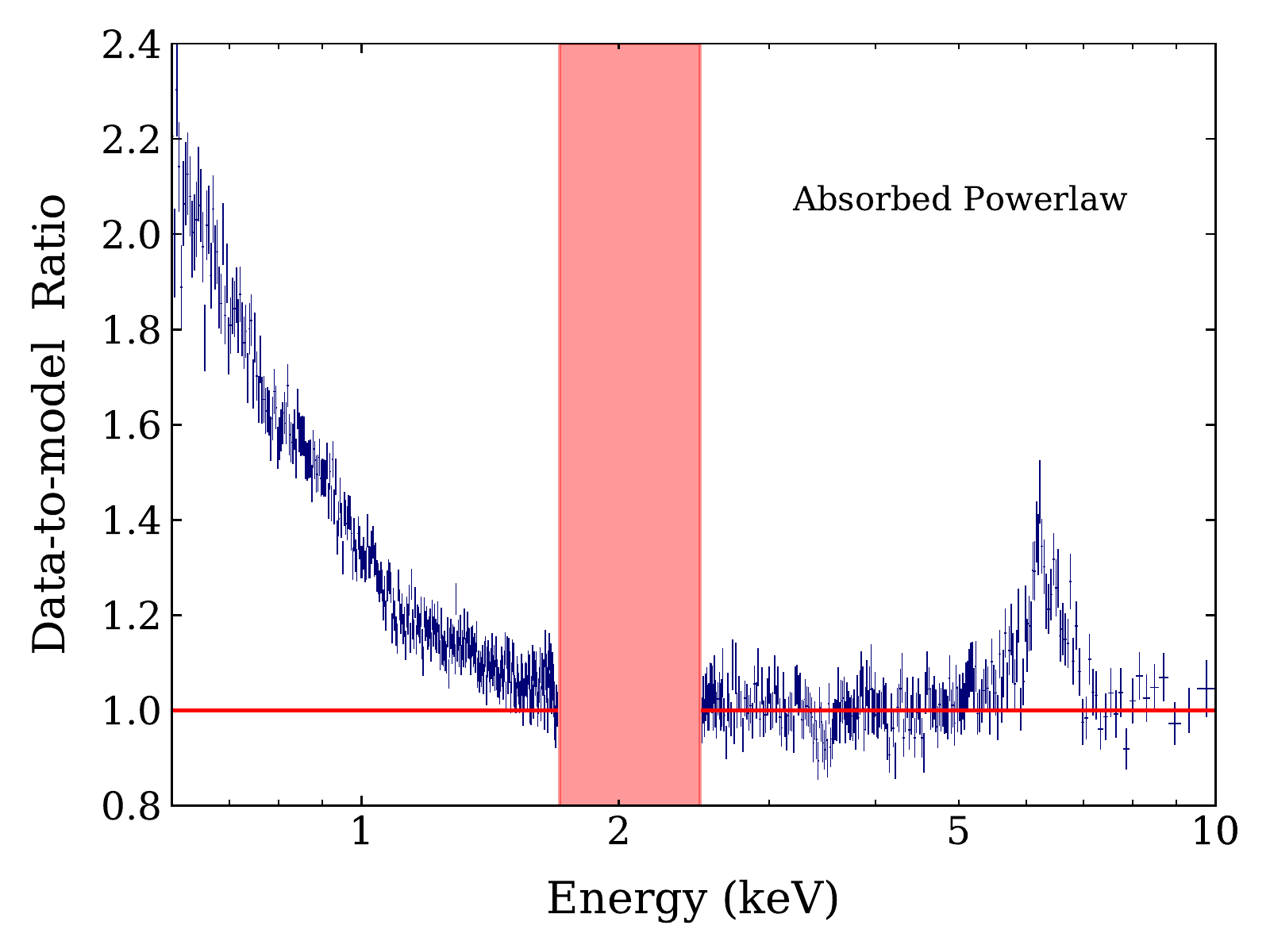}
\caption{A phenomenological powerlaw fit to the time-averaged data with $\Gamma\simeq2.09$. The shaded energy range (1.7--2.5~keV) was not included due to calibration issues. The spectrum was fit from 2.5--4.0 and 8--10~keV, and then plotted after introducing 0.6--1.7 and 4--8~keV bands, to bring out the Fe and soft band features. Plot was rebinned on \textsc{xspec} for clarity.}\label{phenom}
\end{figure}

For our work in this paper we made use of the X-ray spectral fitting package \textsc{xspec} \texttt{v12.10.1b} \citep{XSPEC}. The latest (as of September~4, 2017) ionization balance calculation (\texttt{v3.0.7}) database was imported additionally to be able to properly account for up to date modeling of X-ray emission and absorption from complex spectra.\footnote{\url{http://www.atomdb.org/download.php}}

To bring out features in our 0.6--10~keV spectrum, we first try a phenomenological model-- an absorbed powerlaw fit (Fig.~\ref{phenom}). The galactic H I column has been fixed to $3.56 \times 10^{20}$~cm$^{-2}$ \citep{Kalberla2005} with the ISM grain absorption model \textsc{tbabs}. The cross-sections are set by the \texttt{abund wilm} command \citep{Wilms2000}. The data shows clear signs of a broad Fe K region with a strong soft excess. We ran the fit and obtained poor statistics ($\rchi^2$/d.o.f. = 8011.66/971). The 2--10~keV absorbed powerlaw flux is $\simeq 1.39 \times 10^{-11}$~ergs~cm$^{-2}$~s$^{-1}$. Following what we see, we tested out several model combinations on the XIS dataset. But we only display results from the physically and statistically relevant models, and explain in \S\ref{discuss} why others were not favored. 

For every combination we start with the possibility that only one deformation ($\alpha_{13}$ or $\alpha_{22}$) in the Johannsen metric (See Eqn.~\ref{JM}) is non-zero, leaving it as a fit parameter and assuming all other deformation parameters are equal to 0. For the sake of avoiding unphysically extreme inner disk inclination fits to a Seyfert 1 AGN \citep[e.g.,][]{Nandra1997,Rakshit2017}, we limit the upper hard limit of the inclination parameter in \textsc{relxill\_nk} to $75^\circ$ instead. 
Tables \ref{T1} and \ref{T2} display the best-fit parameters obtained for all models with $\alpha_{13} \neq 0$ and $\alpha_{22} \neq 0$, respectively, while Fig.~\ref{ModelDspec} shows the unfolded spectrum with the contributions of the additive model components, for our best-fit model. 


\subsubsection{Model A} \label{ModA}
\begin{center}
\textsc{tbabs * (zpowerlw + relxill\_nk)}
\end{center}

We added our non-Kerr blurred reflection model \textsc{relxill\_nk} here with the reflection fraction (R$_f$) parameter fixed at -1, corresponding to a pure reflection signal. The redshifted powerlaw component (PLC) takes into account the continuum signal separately, with the photon indices $\Gamma$ linked between the two. Since an E~$<10$~keV dataset  is not suited for constraining the high-energy cutoff ($E_{\textrm{cut}}$) in AGN \citep[e.g., refer to][for typical values]{Tortosa2018}, $E_{\textrm{cut}}$ was fixed at the default value of 300~keV \citep{Keek2016}. Model A $\rchi^2$ plots in Fig.~\ref{R1} and Fig.~\ref{R2} show the immense improvement in the fits, with $\Delta\rchi^2 > 6800$. Most of the soft excess is accounted for by the blurred reflector \citep{Miller2007}. This suggests the need for a reflection-dominated component (RDC). Still, the convergence is poor and suggests the necessity to consider narrow emission residuals (evident from the $\rchi^2$ plots).


\begingroup
\begin{table*}[!t]
\def\arraystretch{1.65}
     \begin{ruledtabular}
     \caption{Best-fit parameter values obtained employing models A to D with only deformation parameter $\alpha_{13} \neq 0$. Errors are given for 90\% confidence, unless explicitly stated. A single emissivity profile was adopted \citep{W13}.}
     \begin{tabular}{ccccccc} \label{T1}
     \multirow{2}*{Component} & \multirow{2}*{Parameter [\textsf{Unit}]} &  \multicolumn{4}{c}{\begin{small}Model Values\footnote{``$(P)$'' against error values implies the parameter has no bound there}\end{small}} \\ \cline{3-6} 
 & & \begin{small}A\end{small} & \begin{small}B\end{small} & \begin{small}C\end{small} & \begin{small}D\end{small}\\ \hline
   
      \multirow{2}*{\textsc{zpcfabs}} & $n_{\textrm{H}}$ [\textsf{10$^{22}$~cm$^{-2}$}] & --& --& --&  $6.57^{+0.88}_{-0.72}$ \\
 											& $CvrFract$ & --& --& --& $0.24^{+0.03}_{-0.04}$ \\ \hline 
											
     \multirow{2}*{\textsc{zpowerlw}} & $PhoIndex$ & $2.35 \pm 0.01$ & $2.35 \pm 0.01$ & $2.35 \pm 0.01$ & $2.50^{+0.02}_{-0.03}$ \\
                                                 & $norm_1$ [\textsf{10$^{-2}$~\begin{footnotesize}ph~cm$^{-2}$~s$^{-1}$~keV$^{-1}$]\end{footnotesize}} & $0.80 \pm 0.01$ & $0.80 \pm 0.01$ & $0.80 \pm 0.01$ & $1.09 \pm 0.06$ \\ \hline
                                                 
     \multirow{7}*{\textsc{relxill\_nk}} & $q$ & $7.08^{+0.42}_{-0.36}$ & $7.36^{+0.93}_{-0.06}$ & $4.11^{+0.39}_{-0.31}$ & $10.00^{~(P)}_{-4.87}$ \\
   												& $a_{*}$ & $0.998^{~~(P)}_{-0.001}$ & $0.998^{~~(P)}_{-0.001}$ & $0.961^{+0.008}_{-0.018}$ &  $0.990^{+0.007}_{-0.035}$ \\
   												& $i$ [\textsf{deg ($^\circ$)}] & $74.2^{~(P)}_{-1.5}$ & $75.0^{~(P)}_{-1.6}$ & $55.8^{+0.9}_{-0.3}$ & $67.4^{+4.1}_{-3.9}$ \\
   												& $log~\xi$ & $1.23^{+0.08}_{-0.11}$ & $1.30^{+0.01}_{-0.14}$ & $1.25 \pm 0.05$ & $1.29^{+0.03}_{-0.17}$ \\
   												& $A_{\textrm{Fe}}$ & $2.30^{+0.32}_{-0.34}$ & $1.61^{+0.44}_{-0.38}$ & $1.39^{+0.20}_{-0.40}$ &  $0.84^{+0.23}_{-0.18}$ \\
   												
   												
   												& $\alpha_{13}$ & $-0.299^{+0.006}_{-0.025}$ & $-0.286^{+0.025}_{-0.018}$ & $-1.195^{+0.121}_{-0.256}$ & $0.214^{+0.087}_{-1.103}$ \\
   												& $norm_2$ [\textsf{10$^{-4}$~\begin{footnotesize}ph~cm$^{-2}$~s$^{-1}$~keV$^{-1}$]\end{footnotesize}} & $3.54 \pm 0.17$ & $3.20^{+0.20}_{-0.26}$ & $2.74^{+0.19}_{-0.10}$ & $2.87^{+0.22}_{-0.33}$ \\ \hline

   \textsc{xillver} & $norm_3$ [\textsf{10$^{-5}$~\begin{footnotesize}ph~cm$^{-2}$~s$^{-1}$~keV$^{-1}$]\end{footnotesize}} & --& $3.33^{+1.10}_{-1.19}$ & $3.45^{+0.88}_{-0.86}$ & $10.93^{+3.12}_{-2.40}$ \\ \hline

   \multirow{3}*{\textsc{zgauss}} & $LineE$ [\textsf{keV}] & --& --& $6.65^{+0.04}_{-0.05}$ & $6.66^{+0.03}_{-0.04}$ \\
   											& $flux$ [\textsf{10$^{-6}$~\begin{footnotesize}ph~cm$^{-2}$~s$^{-1}$]\end{footnotesize}} & --& --& $2.99^{+1.40}_{-1.34}$ & $4.60^{+1.41}_{-1.37}$ \\
   											& EW$_{\textrm{XXV}}$ [\textsf{eV}] & --& --& $18.4^{+9.1}_{-9.0}$ & $28.5^{+9.2}_{-9.4}$ \\ \hline \hline
   			$\rchi^2/d.o.f.$ & -- & 1155.28/964 & 1128.78/963 & 1119.02/961 & 1024.99/959 \\
   			& -- & $\approx$~1.198 & $\approx$~1.172 & $\approx$~1.164 & $\approx$~1.069 \\
   											
       \end{tabular}
       \end{ruledtabular}
\end{table*}
\endgroup

\begingroup
\begin{table*}[!t]
\def\arraystretch{1.65}
     \begin{ruledtabular}
     \caption{Best-fit parameter values obtained employing models A to D with only deformation parameter $\alpha_{22} \neq 0$. Errors are given for 90\% confidence, unless explicitly stated. A single emissivity profile was adopted \citep{W13}.}
     \begin{tabular}{ccccccc} \label{T2}
     \multirow{2}*{Component} & \multirow{2}*{Parameter [\textsf{Unit}]} &  \multicolumn{4}{c}{\begin{small}Model Values\footnote{``$(P)$'' against error values implies the parameter has no bound there}\end{small}} \\ \cline{3-6} 
 & & \begin{small}A\end{small} & \begin{small}B\end{small} & \begin{small}C\end{small} & \begin{small}D\end{small}\\ \hline

      \multirow{2}*{\textsc{zpcfabs}} & $n_{\textrm{H}}$ [\textsf{10$^{22}$~cm$^{-2}$}] & --& --& --&  $6.61^{+0.83}_{-0.72}$ \\
 											& $CvrFract$ & --& --& --& $0.23^{+0.03}_{-0.02}$ \\ \hline
											
     \multirow{2}*{\textsc{zpowerlw}} & $PhoIndex$ & $2.36 \pm 0.01$ & $2.36 \pm 0.01$ & $2.36 \pm 0.01$ & $2.50 \pm 0.02$ \\
                                                 & $norm_1$ [\textsf{10$^{-2}$~\begin{footnotesize}ph~cm$^{-2}$~s$^{-1}$~keV$^{-1}$]\end{footnotesize}} & $0.80 \pm 0.01$ & $0.80 \pm 0.01$ & $0.80 \pm 0.01$ & $1.09^{+0.06}_{-0.05}$ \\ \hline
                                                 
     \multirow{7}*{\textsc{relxill\_nk}} & $q$ & $10.00^{~~(P)}_{-2.12}$ & $10.00^{~~(P)}_{-0.39}$ & $10.00^{~~(P)}_{-0.47}$ & $8.72^{~~(P)}_{-3.82}$ \\
   												& $a_{*}$ & $0.965^{+0.013}_{-0.029}$ & $0.977^{+0.016}_{-0.022}$ & $0.975^{+0.005}_{-0.016}$ &  $0.932^{+0.052}_{-0.046}$ \\
   												& $i$ [\textsf{deg ($^\circ$)}] & $68.2 \pm 1.5$ & $69.9^{+1.45}_{-1.70}$ & $70.3^{+0.8}_{-1.7}$ & $66.8^{+5.6}_{-4.7}$ \\
   												& $log~\xi$ & $1.24^{+0.06}_{-0.10}$ & $1.29^{+0.01}_{-0.15}$ & $1.29^{+0.02}_{-0.14}$ & $1.29^{+0.03}_{-0.17}$ \\
   												& $A_{\textrm{Fe}}$ & $2.16^{+0.32}_{-0.33}$ & $1.45^{+0.41}_{-0.28}$ & $1.18^{+0.48}_{-0.19}$ &  $0.83 \pm 0.16$ \\
   												
   												
   												& $\alpha_{22}$ & $0.010^{+0.089}_{-0.055}$ & $-0.027^{+0.044}_{-0.183}$ & $-0.003^{+0.083}_{-0.073}$ & $0.311^{+0.811}_{-0.613}$ \\
   												& $norm_2$ [\textsf{10$^{-4}$~\begin{footnotesize}ph~cm$^{-2}$~s$^{-1}$~keV$^{-1}$]\end{footnotesize}} & $3.37 \pm 0.18$ & $3.02^{+0.09}_{-0.27}$ & $2.89^{+0.18}_{-0.29}$ & $2.83^{+0.20}_{-0.31}$ \\ \hline

   \textsc{xillver} & $norm_3$ [\textsf{10$^{-5}$~\begin{footnotesize}ph~cm$^{-2}$~s$^{-1}$~keV$^{-1}$]\end{footnotesize}} & --& $4.08^{+1.18}_{-1.11}$ & $4.50^{+1.32}_{-1.19}$ & $11.01^{+2.20}_{-2.50}$ \\ \hline

   \multirow{3}*{\textsc{zgauss}} & $LineE$ [\textsf{keV}] & --& --& $6.65 \pm 0.04$ & $6.66^{+0.03}_{-0.04}$ \\
   											& $flux$ [\textsf{10$^{-6}$~\begin{footnotesize}ph~cm$^{-2}$~s$^{-1}$]\end{footnotesize}} & --& --& $2.99^{+1.40}_{-1.37}$ & $4.60^{+1.42}_{-1.43}$ \\
   											& EW$_{\textrm{XXV}}$ [\textsf{eV}] & --& --& $18.4 \pm 9.0$ & $28.5^{+9.2}_{-9.1}$ \\ \hline \hline
   			$\rchi^2/d.o.f.$ & -- & 1188.04/964 & 1145.12/963 & 1133.14/961 & 1024.90/959 \\
   			& -- & $\approx$~1.232 & $\approx$~1.189 & $\approx$~1.179 & $\approx$~1.069 \\
   											
       \end{tabular}
       \end{ruledtabular}
\end{table*}
\endgroup

\subsubsection{Model B} \label{ModB}
\begin{center}
\textsc{tbabs * (zpowerlw + relxill\_nk + xillver)}
\end{center}

We made use of the \textsc{xillver} reflection code here to account for narrow, non-relativistic reflected emission in the spectrum. From preliminary fits we find that leaving ionization parameter free here leads to fits of $log~\xi_{\textrm{unblurred}} \approx 0$. Similarly, the fit is insensitive to the value of the inclination parameter of the distant reflector. Hence, we fixed ionization to 0 and inclination to the default. The iron abundance $A_{\textrm{Fe}}$ was linked to that of the blurred reflector with the general idea that the cold reflector could well be part of the same galaxy. Leaving $A_{\textrm{Fe}}$ fixed at solar value returns poor fits. On the other hand, leaving it free results in extremely high, unconstrained abundances at the upper bound. The addition is $>4\sigma$ significant for $\alpha_{13}$ ($\Delta\rchi^2 \simeq 26$) and $>5.5\sigma$ significant for $\alpha_{22}$ ($\Delta\rchi^2 \simeq 43$), for 1 extra degree of freedom (d.o.f.). Model B $\rchi^2$ plots have not been shown because they are very similar to Model C's. (See below.) The middle panel in figures \ref{R1} and \ref{R2} clearly shows that the narrow Fe core was fit by the model.

\subsubsection{Model C} \label{ModC}
\begin{center}
\textsc{tbabs * (zpowerlw + relxill\_nk + xillver + zgauss)}
\end{center}

We detected the presence of a narrow ($\sigma=10$~eV) Fe-\textrm{XXV} emission line at E~$\simeq 6.65$~keV for both $\alpha_{13}$ and $\alpha_{22}$ cases. Existence of this residual was also mentioned in \cite{Patrick2011} and \cite{W13}, who analyzed the same dataset in the Kerr spacetime. The improvement in delta-fit statistic with the inclusion of such a line in the model agrees with \cite{Patrick2011} (Tab.~8 therein) at $\Delta\rchi^2 \simeq 10$.

\subsubsection{Model D} \label{ModD}
\begin{center}
\textsc{tbabs * [zpcfabs * (zpowerlw + relxill\_nk + xillver + zgauss)]}
\end{center}

\begin{figure}[!t]
\centering
\includegraphics[width=0.475\textwidth, trim={4mm 4mm 4mm 2mm},clip]{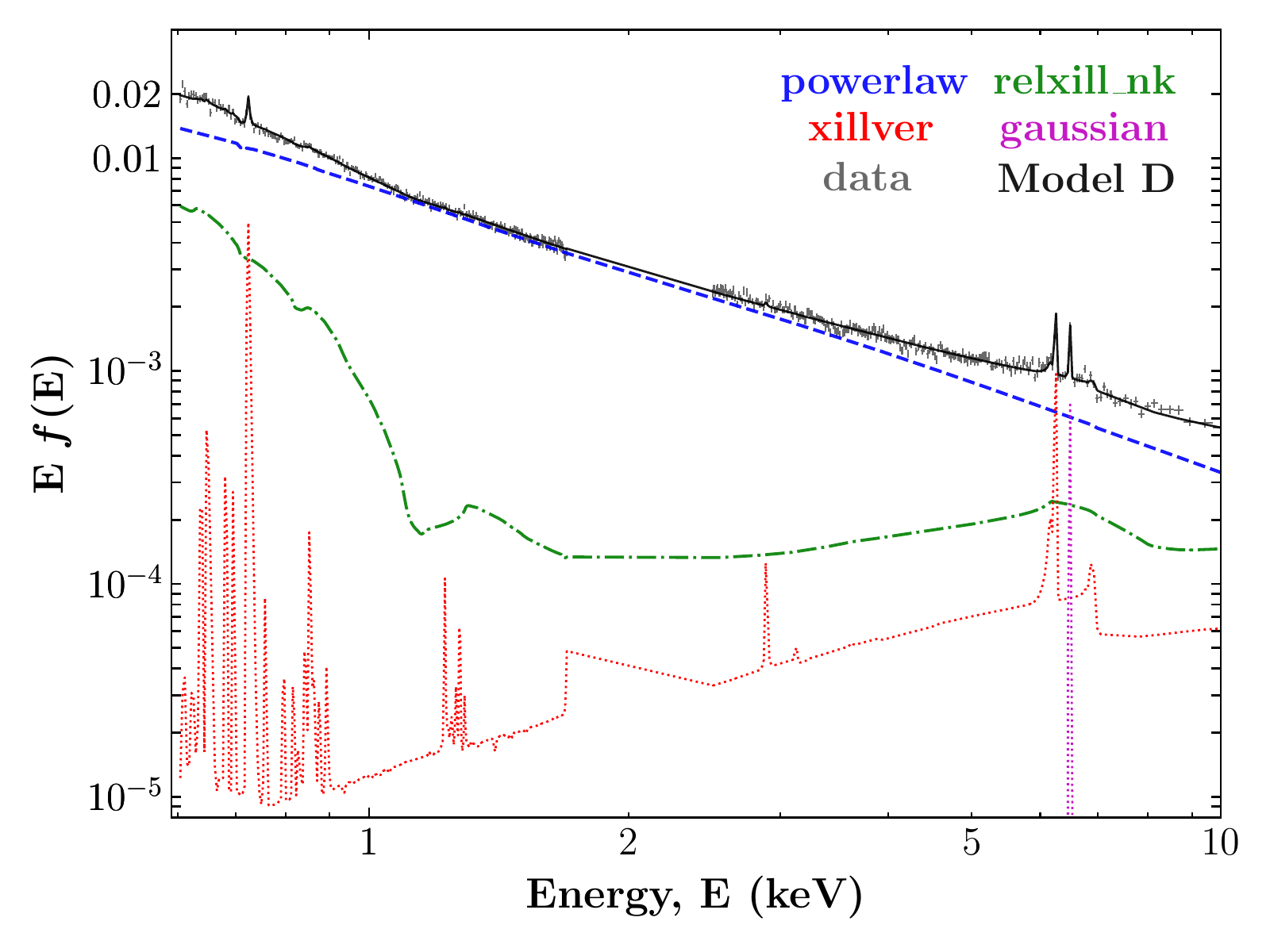}
\caption{Plot shows the unfolded spectrum with the best-fit model D. The vertical axis is in units of photons~cm$^{-2}$~s$^{-1}$. Contributions from different model components are shown in different colors.} \label{ModelDspec}
\end{figure}

After Model C, we tried several alternative RDC combinations, briefly described in \S\ref{discuss}. But none of them seemed to improve the fit statistic. However, hints of absorption can be seen in the spectra. We turned to this possibility, and different absorbers and their combinations were examined. The simplest inclusion of one partial covering (\textsc{zpcfabs}) proved satisfactory. The $\Delta\rchi^2$ in the $\alpha_{13}$ and $\alpha_{22}$ cases were, respectively, 94 and 108 for 2 extra d.o.f.\,. The parameters of the covering model are well-constrained as can be seen in Tab.~\ref{T1} and Tab.~\ref{T2}. The inclusion proved to be of high significance ($>7\sigma$) in improving the fits. The unfolded spectrum of model D is plotted Fig.~\ref{ModelDspec}, along with the contribution of each additive model component.

\begin{figure*}[!t]
\begin{center}
        \subfigure[]{

            \includegraphics[width=0.47\textwidth, trim={2mm 0 0 1.2cm},clip]{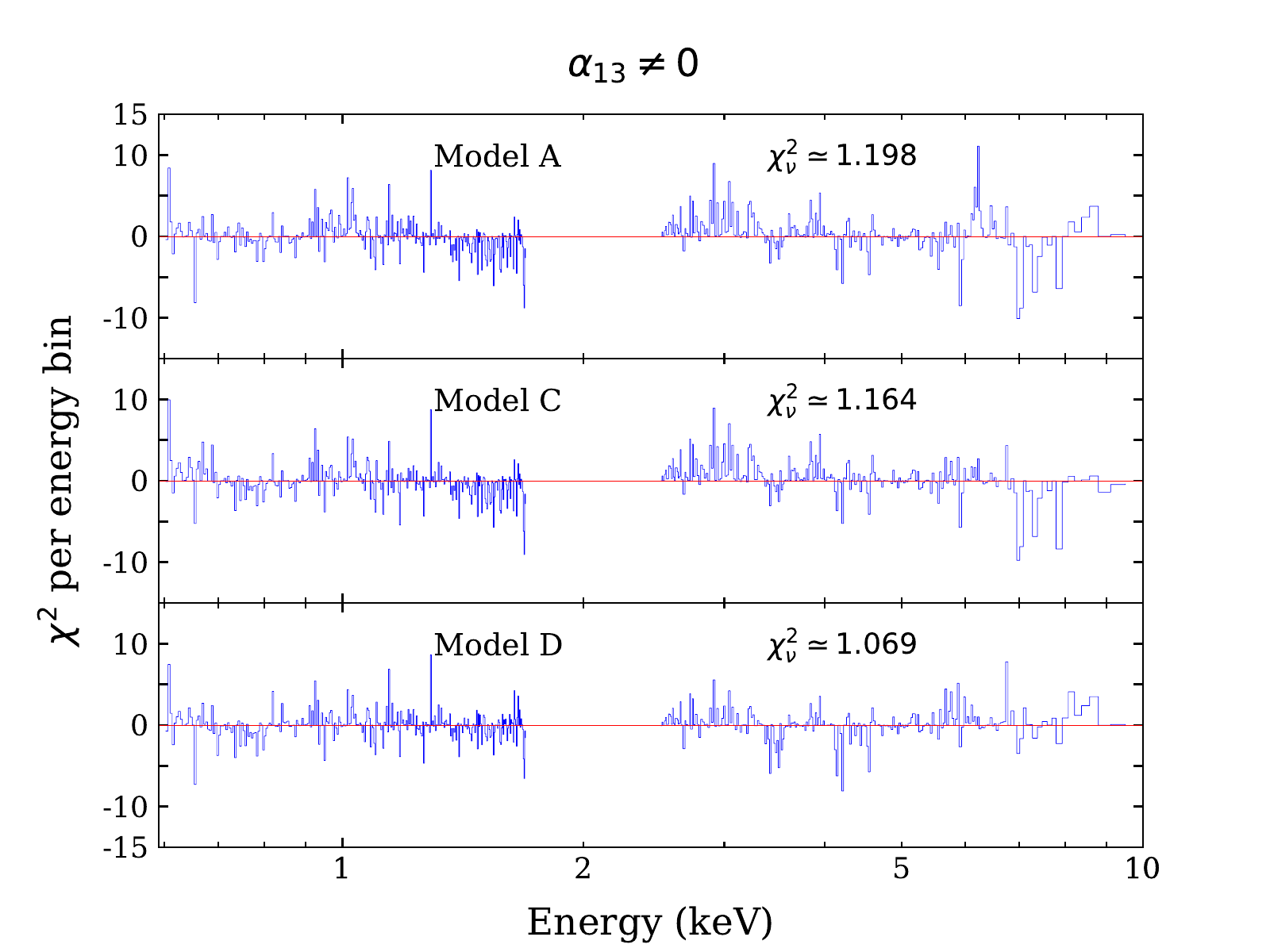}\label{R1}
        }
        \quad
        \subfigure[]{
	\includegraphics[width=0.47\textwidth, trim={2mm 0 0 1.2cm},clip]{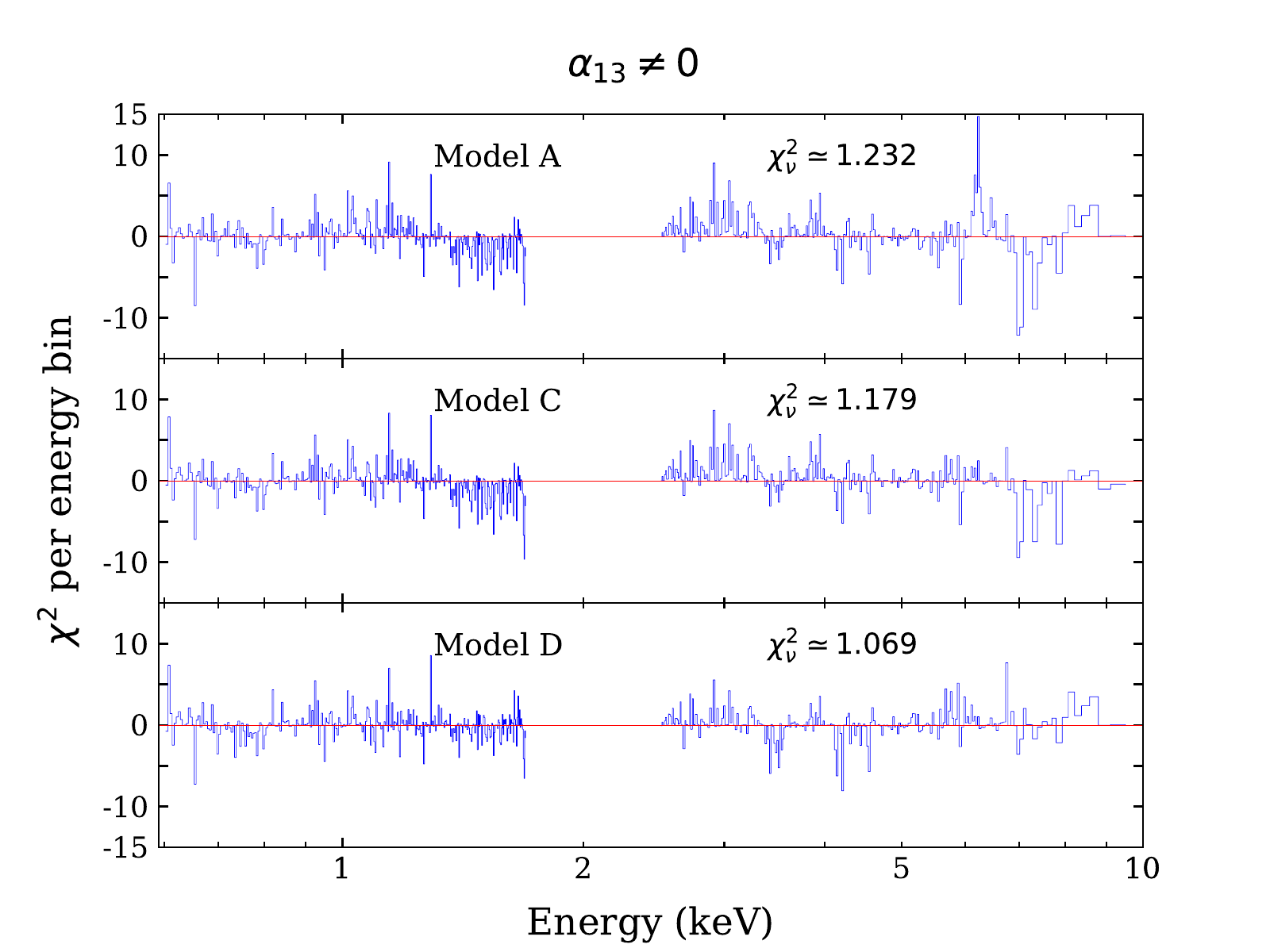}\label{R2}
           
        }
        
     \caption{$\rchi^2$ plots for models A, C and D for $\alpha_{13} \neq 0$ (left) and $\alpha_{22} \neq 0$ (right) from Tab.~\ref{T1} and Tab.~\ref{T2}, respectively. Plots were rebinned on \textsc{xspec}. $\nu$ in $\rchi^2_{\nu}$ represents the degrees of freedom (d.o.f.), where $\rchi^2_{\nu} = \rchi^2/d.o.f.$} 
     \label{D1}
\end{center}
\end{figure*}

\begin{figure*}[!t]
\begin{center}
        \subfigure[]{
           \includegraphics[width=0.475\textwidth, trim={3.4cm 2.8cm 3.3cm 2.2cm},clip]{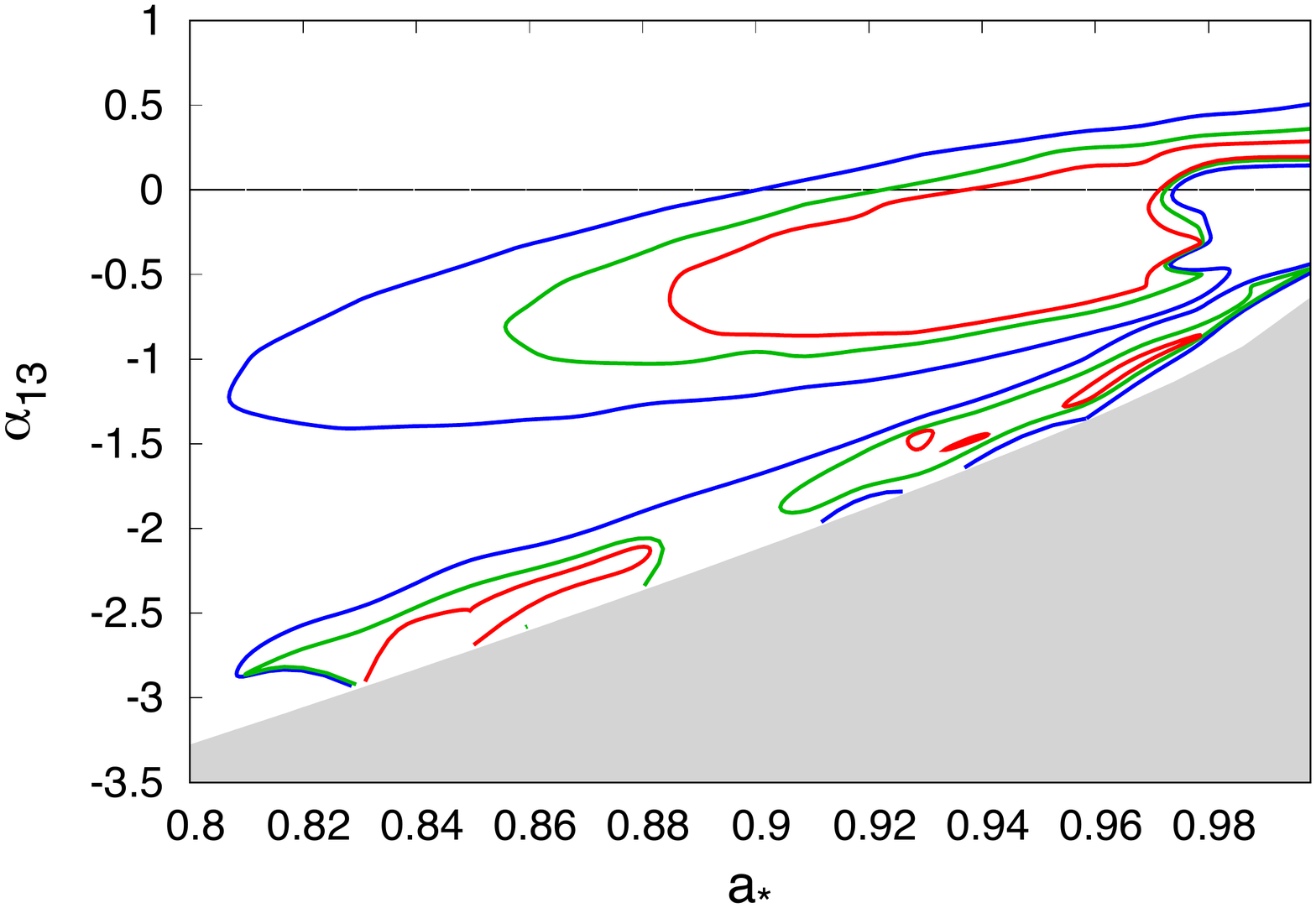} \label{A13Z}
        }
        \quad
        \subfigure[]{
            \includegraphics[width=0.475\textwidth, trim={3.4cm 2.8cm 3.3cm 2.2cm},clip]{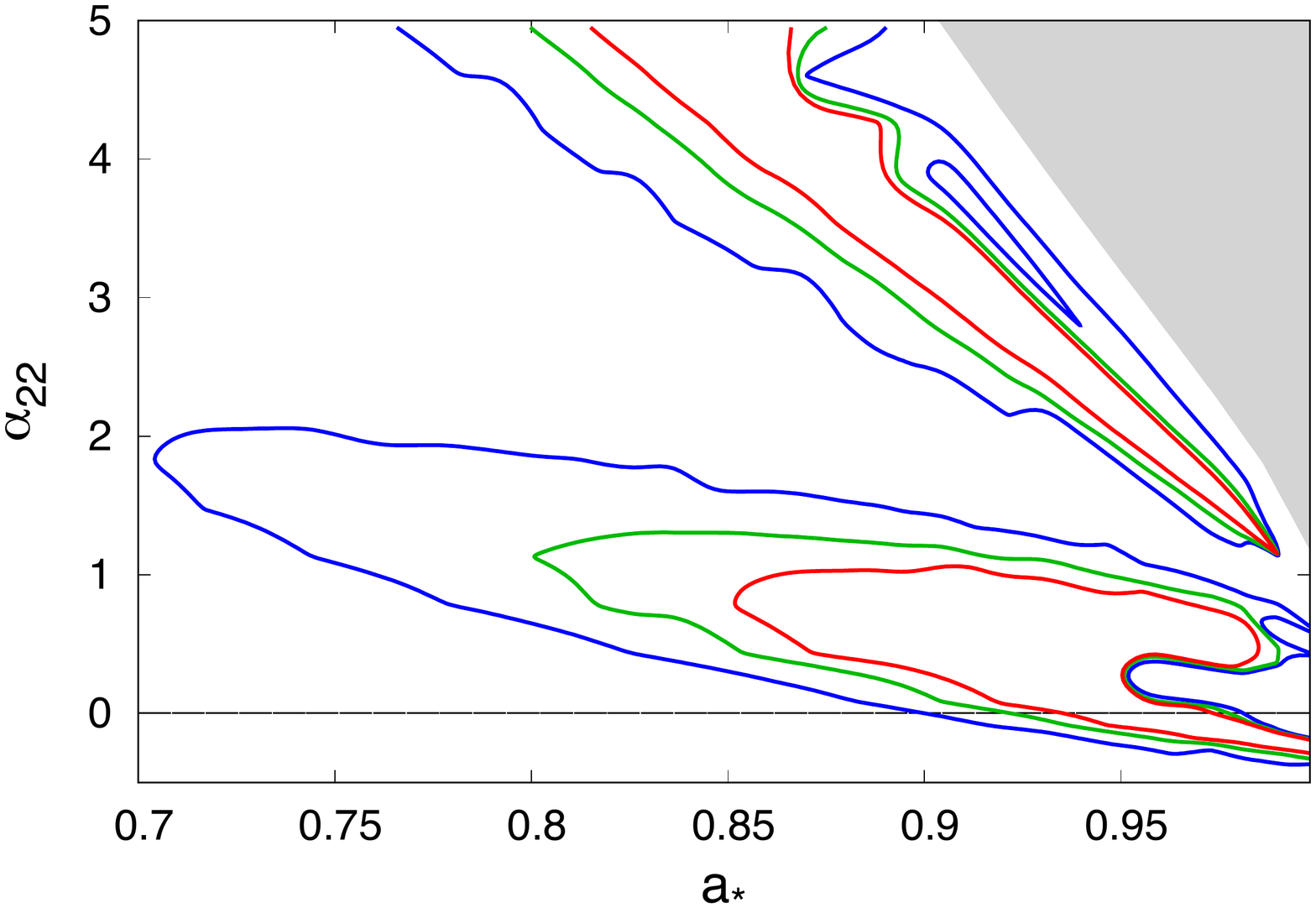}\label{A22Z}
        }
        
     \caption{Two-dimensional degeneracy contours between the dimensionless spin parameter $a_*$ and Johannsen deformation parameters $\alpha_{13}$ (left) and $\alpha_{22}$ (right) for the best-fit model D. The dashed, horizontal line at 0 marks the Kerr solution. The red, green and blue contour lines indicate 68\%, 90\% and 99\% confidence levels, respectively. The shaded spaces in gray are forbidden regions within the metric that avoid some pathological properties as mentioned in \S\ref{metric}.} \label{ALPHA}
\end{center}
\end{figure*}

\section{Results} \label{sec:results}

The primary aim of this study is to provide constraints on deviations from the Kerr metric. Tab.~\ref{T1} and Tab.~\ref{T2} list the values of $\alpha_{13}$ and $\alpha_{22}$ for each model. For the best-fit model, model D, the Kerr solution (corresponding to deformation parameter equal to zero) is recovered in both cases. Since deformation parameters are strongly degenerate with the spin parameter, we show the two-dimensional spin-deformation contours in Fig.~\ref{ALPHA} for $\alpha_{13}$ and $\alpha_{22}$, using the \textsc{steppar} command on \textsc{xspec}. There is a clear degeneracy between the two parameters, which includes the Kerr solution in both cases. Apart from the usual contours observed in such studies (the \emph{near-Kerr} contours), cf. Fig. 2 in~\cite{TripathiArk564} and Fig. 2 in~\cite{Xu2018}, additional contours appear which exclude the Kerr solution at $3\sigma$ (the \emph{far-Kerr} contours). A thorough investigation (detailed in~Appendix~\ref{app:solutions}) indicates that these \emph{far-Kerr} contours are unreliable. We thus ascribe no significance to these. The constraints on the deformation parameters from these plots, after excluding the \emph{far-Kerr} contours, are:
\begin{align}
\alpha_{13} &\neq 0: \hspace{7mm} a_* > 0.8      &-1.5 < \alpha_{13} < 0.6 \nonumber \\
\alpha_{22} &\neq 0: \hspace{7mm} a_* > 0.7    & \ \ \ -0.4<\alpha_{22}<2.1
\end{align} \label{constraints}
We can make some comparisons for $\alpha_{13}$ with constraints from other studies. The constraints obtained here are better than those in \citealt{Bambi2018} (Fig.~2 and Fig.~3 therein), which are the updated versions of the same in \cite{Cao2018} for both 2011 XMM-Newton data and simultaneous \textit{NuSTAR}+\textit{Swift} results of the NLS1 1H0707--495. The contours with Markov Chain Monte Carlo (MCMC) runs in \cite{Jingyi2018} for the popular X-ray BHB GX~339--4 with \textit{RXTE} PCU-2 data are comparable to ours here. For both deformation parameters, results from \cite{TripathiArk564} on the NLS1 Ark~564 \textit{Suzaku} XIS and from \cite{Xu2018} on the low-mass X-ray binary GS~1354--645 \textit{NuSTAR} FPM analyses are stronger than ours here. 

Note that one of systematic effects that affects the uncertainty in deformation parameter values is the ISCO (innermost stable circular orbit) radius degeneracy between spin and the deformation parameter. (See, e.g., Fig.~6 in~\cite{Johannsen2013} and Fig.~12 in~\cite{Tripathi2019}.) Lower spins result in weaker constraints on the deformation parameters. We can see this pattern in all the results: For GX~339--4, the spin is comparable to the spin here, so is the $\alpha_{13}$ uncertainty; for Ark~564 and GS~1354--645, the spin is higher and is more strongly constrained, and so $\alpha_{13}$ also has stronger constraints. 

The spin recovered is consistent with literature where Mrk~335 can be seen to have wide uncertainties on $a_*$ in the Kerr background at 90\% confidence for one parameter of interest \citep[e.g., $a_*>0.7$ in][with the $\sim$200~ks \textit{XMM Newton} intermediate-flux data from 2009]{Gallo2013}. Even \cite{W13} presented a lower limit of $0.7$ on $a_*$ with our very dataset. Our best-fit $a_*$ from Tab.~\ref{T1} and Tab.~\ref{T2} are similar to those from \cite{Gallo2015}, who use a broken emissivity profile instead. We roughly recover $0.93 \lesssim a_* \lesssim 0.97$ in the Kerr limit ($\alpha_{13}=\alpha_{22}=0$) from both Fig.~\ref{A13Z} and Fig.~\ref{A22Z}, at 90\% confidence. 



\section{Discussion} \label{discuss}

We now discuss some aspects of the models and the results. 
\subsection{Soft Excess}
Absorption in the $1-3$~keV band is evident from absorbed powerlaw fit, with a strong soft excess $<1$~keV, both being reported in \cite{Keek2016}. From model fits A to C we see tight parameter constraints, possibly indicating overestimation of error bars. Comparing with \cite{Patrick2011} and \cite{W13}, we observe a rise in $\Gamma$ and drop in $\xi$ with all four models. This line-continuum trade-off is possible with \textsc{relxill} \citep{K17}. Recovering a relatively colder disk can also be responsible for causing the soft excess at lower energies, resulting from blending of multiple narrow emission lines that can be fit by blurred reflection \citep{Miller2007}. 

\subsection{Inclination Angle}

\begin{figure}[!t]
\centering
\includegraphics[width=0.475\textwidth, trim={1.9cm 7mm 4cm 1.9cm},clip]{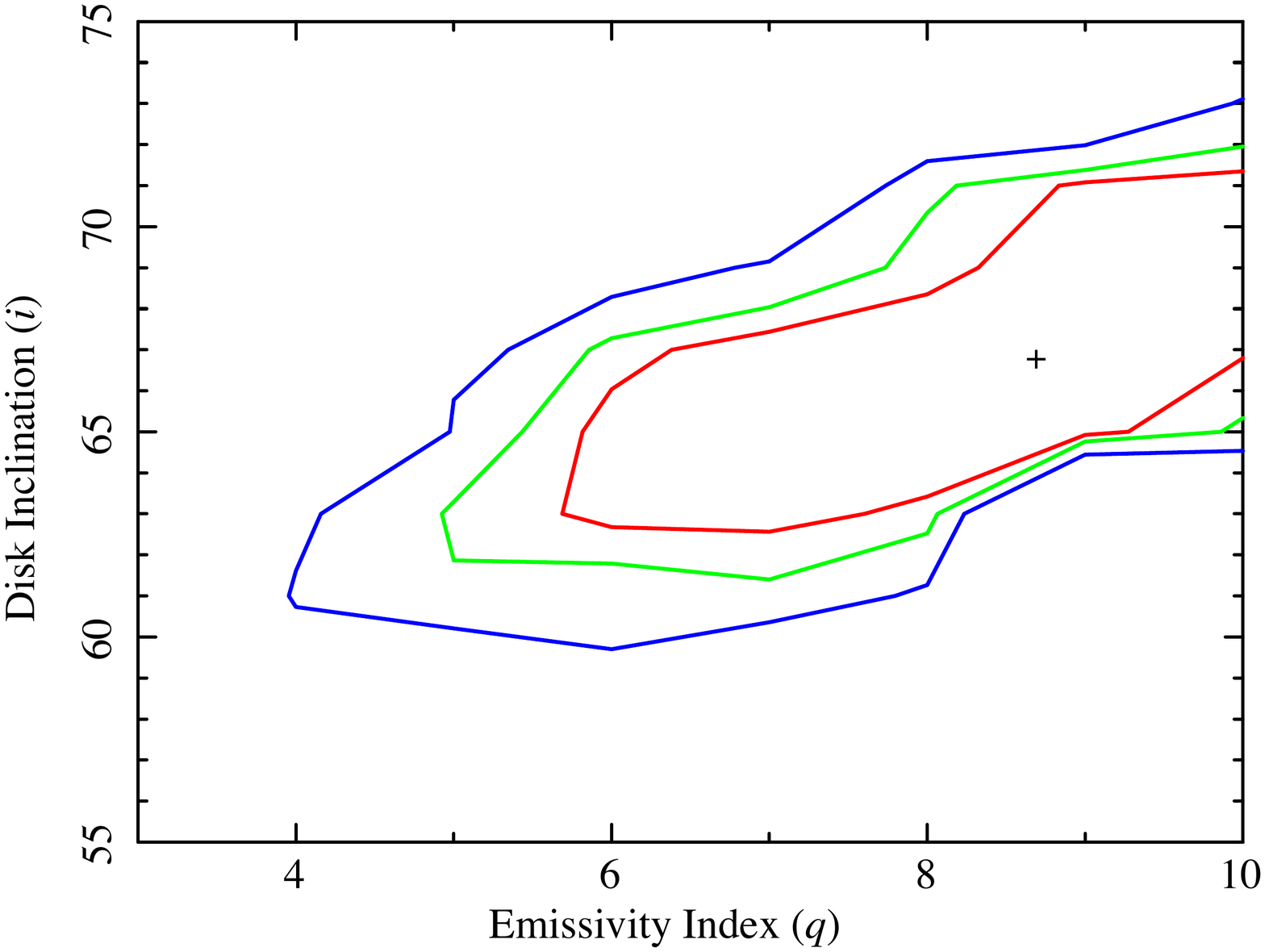}
\caption{Contour plot between the inner disk inclination $i$ (in units of degrees) and emissivity index $q$ from model D, $\alpha_{22}\neq0$ case. The red, green and blue contour lines indicate 68\%, 90\% and 99\% confidence levels, respectively. The best-fit is marked with a ``\textbf{+}''.}
\label{i-q}
\end{figure}

In general, a Seyfert 1 AGN (especially a NLS1) is seen more face-on than edge-on, while we obtain a best-fit $i \simeq 67^\circ$. This is comparatively higher than what \cite{W13} report at similar reduced statistic, although we have higher precision. One possible explanation is as follows: the fit values in both \cite{Patrick2011} and \cite{W13} seem to follow the $i-q$ degeneracy trend shown in \cite{Parker2014} (Fig.~11 therein) for time-averaged spectrum analyzed with \textsc{relxill}. We also find this degeneracy (See Fig.~\ref{i-q}) with a preference for $i > 58^\circ$ at 99\% confidence.
Owing to limited spectral resolution, there could exist a preference for both parameters to yield high fit values in such case. Another possible explanation is that the broad-line region of the AGN is either absent or hidden from the observer which makes us recover a higher angle \citep{Giannuzzo1996}. 

\subsection{Absorption}
Although Model C here is analogous to the model in \citealt{W13} (who employ \textsc{reflionx} instead; \citealt{RF2005}), we see that the fits are not satisfactory. We attempted to fit a doubly-blurred reflection model (a second \textsc{relxill\_nk}) with all parameters linked, except for the ionization and normalization \citep{Cao2018,TripathiArk564}, both with and without the distant reflector. We also inspected with multiple distant reflectors. The inclusions did not improve the fits, implying there may be no accretion disk inhomogeneities as such. A general (or multi-component) PLC+RDC model combination cannot account for this dataset with \textsc{relxill\_nk}. The spectra still had significant residuals beyond $2\sigma$ significance at energies $<3$~keV, which contains $\sim80\%$ of the total XIS counts. 

To obtain a better fit, we looked at the $\rchi^2$ plots. (See model C in~\ref{R1} and~\ref{R2}.) Absorption within the observed (redshifted source) frame E~$\sim7-8$~keV can be seen. Local absorption at the source frame seems possible as mentioned in \cite{Longinotti2013} and \cite{Keek2016} (See \citealt{Larsson2008} and \citealt{Patrick2011} for contrary opinions). We decided to add absorbers to the model and found a better fit. While both ionized and unionized absorbers provided better fits, eventually the unionized absorber model \textsc{zpcfabs} was found to be satisfactory. (See Appendix~\ref{app:absorb} for details.)



We now discuss the implications of the introduction of such an absorption model in the analysis. Even in the Kerr case, there are results in the literature where we see that a mild continuum absorption of the reflection data improves the fits. Very recent work on the NLS1 Mrk~766 shows that this hybrid scenario is possible and yields more reasonable estimates on the X-ray continuum absorption \citep{Buisson2018}. The referred work finds parameter constraints in-line with ours after introducing a partial covering to their reflection model. One of the effects of such absorption is the increase in uncertainty on some or all parameters of interest. We can see such an effect on the estimation of the deformation parameters between Model D and all other models (see, Tab.~\ref{T1} and Tab.~\ref{T2}). 

Model D, which includes the absorber, gives us a similar converging best-fit statistic ($\sim$1.07) as the best-fit model in \cite{W13} does. Moreover, in Model D both the Fe-\textrm{XXV} line energy and the equivalent width (EW$_{\textrm{XXV}}$) of the line are similar to that reported in \cite{W13}, after possible broadening introduced due to the adjustment offered by the partial covering, reducing the need for narrow emission lines. Additionally, we see in Model D enhanced signal (rise in the intensity of distant reflector from $\sim$12\% in Model C to $\sim$28\% in Model D compared to the total RDC contribution; a rise in $\Gamma$ by $\sim$6\%) as a compensation for the flux absorbed, possibly from both the continuum and the cold reflector. At high $i\simeq67^\circ$, the partial covering barely affects the net model complexity with a Compton-thin continuum absorption under a low $\sim$25\% covering, pointing at possible line-of-sight obstruction by local (cold) matter. The net X-ray flux absorbed at the source rest frame from Model C to D is also low ($\Delta L_{\textsf{0.6-10~\textrm{keV}}}\simeq 1-1.5 \times 10^{-2}$~ergs~s$^{-1}$).

\subsection{Caveats}
Our constraints from this work establish that the Kerr metric in Einstein's gravity is recovered at 99\% confidence with the data analyzed. However, care needs to be taken while interpreting the results as there are simplifications involved. The caveats can be divided into four categories: 

\begin{enumerate}
\item There could exist inter-component degeneracies since we include multiple components, and this may add to the wide uncertainties seen in our results. E.g., this is the first time that a partial covering of the reflection spectrum has been proposed for this source in a high-flux dataset. Generically, partial absorption and relativistic emission are degenerate with respect to the spectral shape, and a more sophisticated analysis like MCMC is required to check for robustness of such a combination. 

\item The assumption that the intensity profile is a power law is also an approximation. Employing \textsc{reflionx}, \cite{Gallo2015} and \cite{WG2015} show that the Mrk~335 corona during the \textit{Suzaku} 2006 observation may have been extended, with a collimated jet-like emission away from the accretion disk up to $\sim 26$ gravitational radii (R$_\textrm{g}$) above the back hole's rotational axis and confined to a small region within a few R$_\textrm{g}$. They argue that particles were accelerated away from the accretion disk with the base of the unresolved jet possibly moving non-relativistically, and the particles would return under the gravitational effects of the central engine to still sufficiently illuminate the disk with a low ($< 1.0$) reflection fraction yield. 

\item Even though we have sampled high counts per bin with $\rchi^2$-statistics here, we can still be subject to bias in results since our total number of data bins to be fitted (973) is not lower than the squared-root of the total number of counts in the analyzed energy range \citep{cstat2009}.

\item The treatment of X-radiation from astrophysical black holes assuming the thin disk model is computationally and physically stable at large, but constitutes typical, oversimplifying assumptions on the disk conditions \citep{Page1974,Bambi2017book}. \cite{Taylor2018} shows a recent work using the lamppost geometry for relativistic reflection spectroscopy using a thick accretion disk. It may be likely that the disk may thicken (geometrically) with increasing mass accretion rate. At Eddington luminosity $>10$\%, the inner edge of the accretion disk may move further inward than in the thin disk approximation where it is thought to be at the ISCO \citep{AbraLasota1980}. \cite{Keek2016} find the analyzed dataset here to be on this high end of accretion rate with an extended corona, which seems to agree with the coronal geometry proposed in \cite{Gallo2015,WG2015}. Work is currently underway to model and test the effects with a thick disk. 
\end{enumerate} 


\section{Conclusions} \label{conc}

The main goal of this work is to test how well we can recover the Kerr metric if we introduce a deformation in the Kerr spacetime around the supermassive black hole of NLS1 Mrk~335, using observational data. We use the \textsc{xspec} model \textsc{relxill\_nk}, which is an extension of the widely-used relativistic reflection code \textsc{relxill}, to include parametrically-deformed Kerr metrics. We use the \textit{Suzaku} FI XIS dataset, studied in a large survey paper~\cite{W13}. 

We find differences from the previous work of \cite{W13}. A PLC+RDC model does not explain the data completely and partial covering is needed to account for absorption. 
We obtain decent constraints on $a_*$, $\alpha_{13}$ and $\alpha_{22}$ (shown in Fig.~\ref{ALPHA}). At first look, the results indicate two sets of contours, the \emph{near-Kerr} contours including the Kerr solution and following the results from previous studies, and the \emph{far-Kerr} contours excluding the Kerr solution and including extreme values of deformation parameters. A thorough investigation (See Appendix~\ref{app:solutions}) suggests that the \emph{far-Kerr} contours are not robust and sometimes have unreasonable parameter values. Their appearance in the planes of the deformation parameters against $a_*$ and absence with respect to other physical parameters like $log~\xi$ and $i$ with repeated tests indicates there exist uncertainties (possibly data quality and systematics snarled together) currently beyond our control. This is one reason to also say that it is not possible to completely rule out the \emph{far-Kerr} solutions because they exist in at least one two-dimensional parameter phase space. However, except for the spin-deformation contours, all other tests undoubtedly vote in favor of the \emph{near-Kerr} solutions, and thus, our inference in \S\ref{sec:results}.




The data analyzed in this study was found to have a best-fit output $\sim14-15$\% relative strength of relativistic reflection compared to the total contributed by the PLC and the blurred reflector in the RDC.\footnote{This was determined with the \textsc{cflux} model on \textsc{xspec} in the 0.6--10~keV band.} This is the lowest reflection dataset of Mrk~335 from Tab.~\ref{T-obs}. The proof of this also comes from \cite{WG2015} who proposed a change in coronal geometry from the 2006 \textit{XMM Newton} state to the \textit{Suzaku} state (later the same year), to have the become rather extended and beaming more radiation away from the disk in the latter state. The current study, seen from a different perspective, tests for a lower limit with the method of constraining gravity with weak reflection data for Mrk~335. The bounds can only get better from here with a synergy between stronger reflection, better photon counts and appreciable, simultaneous high-energy coverage (See \S\ref{Obsr.} for a recap on available data and states). 

Precision test of gravity using X-ray reflection spectroscopy is a nascent field and every study opens new challenges. Mrk~335 is not an easy source to analyze. The model employed here is the first time \textsc{relxill\_nk} is combined with a partial covering absorber. The analysis performed here had unique challenges, and the fact that we could obtain constraints comparable to other studies suggests that it is indeed possible to use this technique for testing theories of gravity even in complex scenarios, provided one is careful during the analysis and interpretation of results. The constraints presented in this work are not the strongest with the model (compared to other works in recent past) owing to the reflection data quality, and certainly does not display limitations of the X-ray reflection spectroscopy method.


\acknowledgments
We extend our gratitude to the anonymous referees for suggesting helpful tips in order to improve the quality of the paper. K.C. thanks Michael Parker and Matteo Guainazzi for their helpful comments on the data analysis part. The work of K.C., D.A., A.B.A. and C.B. was supported by the National Natural Science Foundation of China (NSFC), Grant No. U1531117, and Fudan University, Grant No. IDH1512060. K.C. also acknowledges support from the China Scholarship Council (CSC), Grant No. 2015GXYD34. S.N. acknowledges support from the Excellence Initiative at Eberhard-Karls Universit{\"a}t T{\"u}bingen. A.B.A. also acknowledges the support from the Shanghai Government Scholarship (SGS). C.B.and J.A.G. also acknowledge support from the Alexander~von~Humboldt Foundation.


\appendix

\section{The near-Kerr and far-Kerr contours} \label{app:solutions}
We now discuss the \emph{near-Kerr} and \emph{far-Kerr} contours of Fig.~\ref{ALPHA}. We can see that both panels indicate the presence of multiple minima bound at 99\% confidence. While the \emph{near-Kerr} contours remain consistent in shape and size when the stepping resolution (i.e., spacing between two values of a parameter in the grid selected) is changed, the \emph{far-Kerr} contours change significantly with changing resolution. Moreover, often the fits landing in the \emph{far-Kerr} region tend to get stuck, and fail to constrain the deformation parameter at both upper and lower bounds. We emphasize at this point that the \emph{near-Kerr} and \emph{far-Kerr} solutions in each panel of Fig.~\ref{ALPHA} are statistically similar (i.e., they have similar $\rchi^2/d.o.f.$). Even best-fit values of parameters (other than $a_*$, $\alpha_{13}$ and $\alpha_{22}$) show no systematic difference between fits in the two solution regions. Since the \emph{far-Kerr} feature has never been observed in other similar analysis, we need to check if both \emph{near-Kerr} and \emph{far-Kerr} contours are physically reasonable and robust. Only statistical similarity is not sufficient to reach a conclusion when conducting sensitive tests, like with strong gravity, when a lot of parameters are involved in a complex modeling.

\begin{figure}[t]
\begin{center}
        \subfigure[]{
           \includegraphics[width=0.5\textwidth, trim={3.4cm 1.8cm 0 2.2cm},clip]{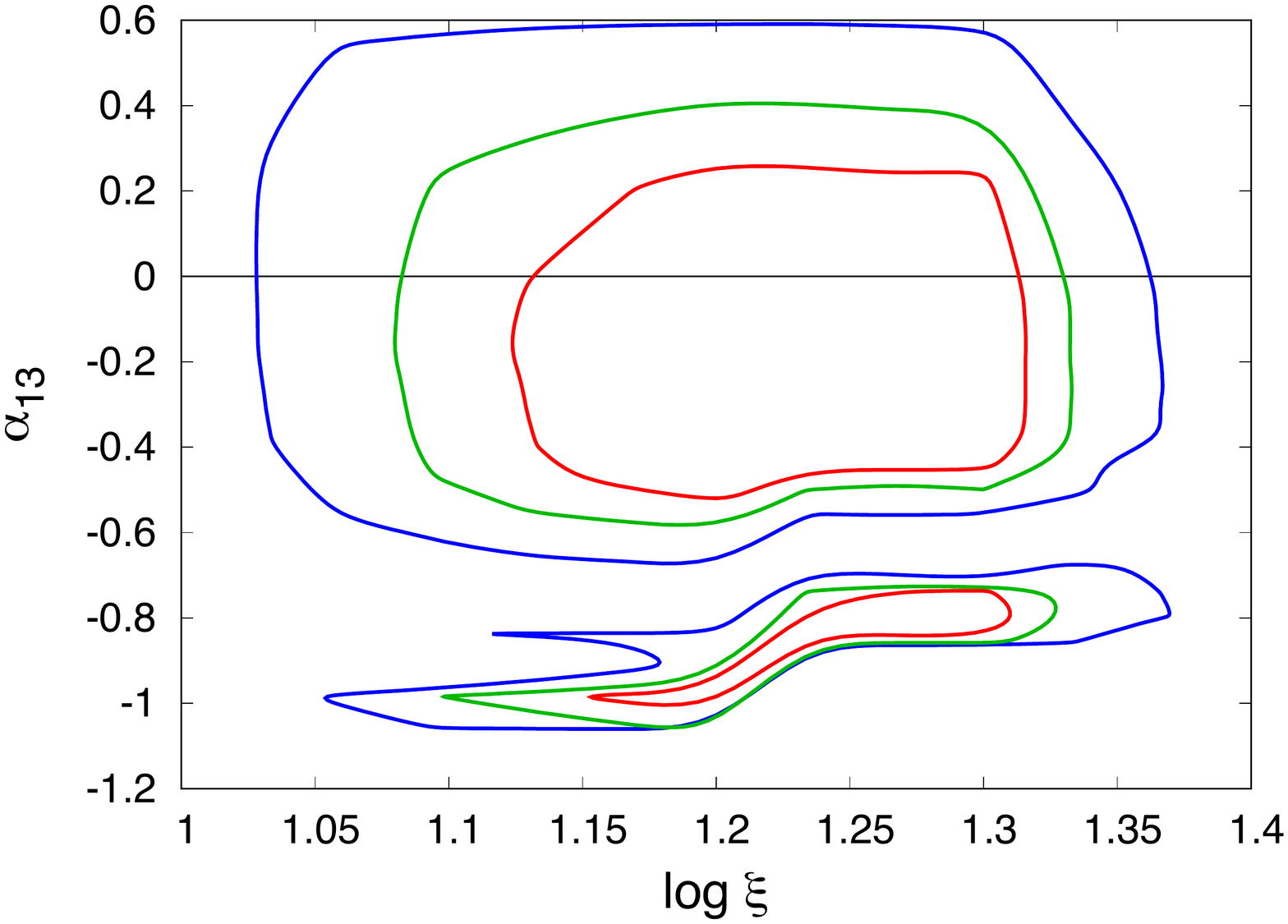} \label{logxi}
        }
        \quad
        \subfigure[]{
            \includegraphics[width=0.5\textwidth, trim={3.4cm 1.8cm 0 2.2cm},clip]{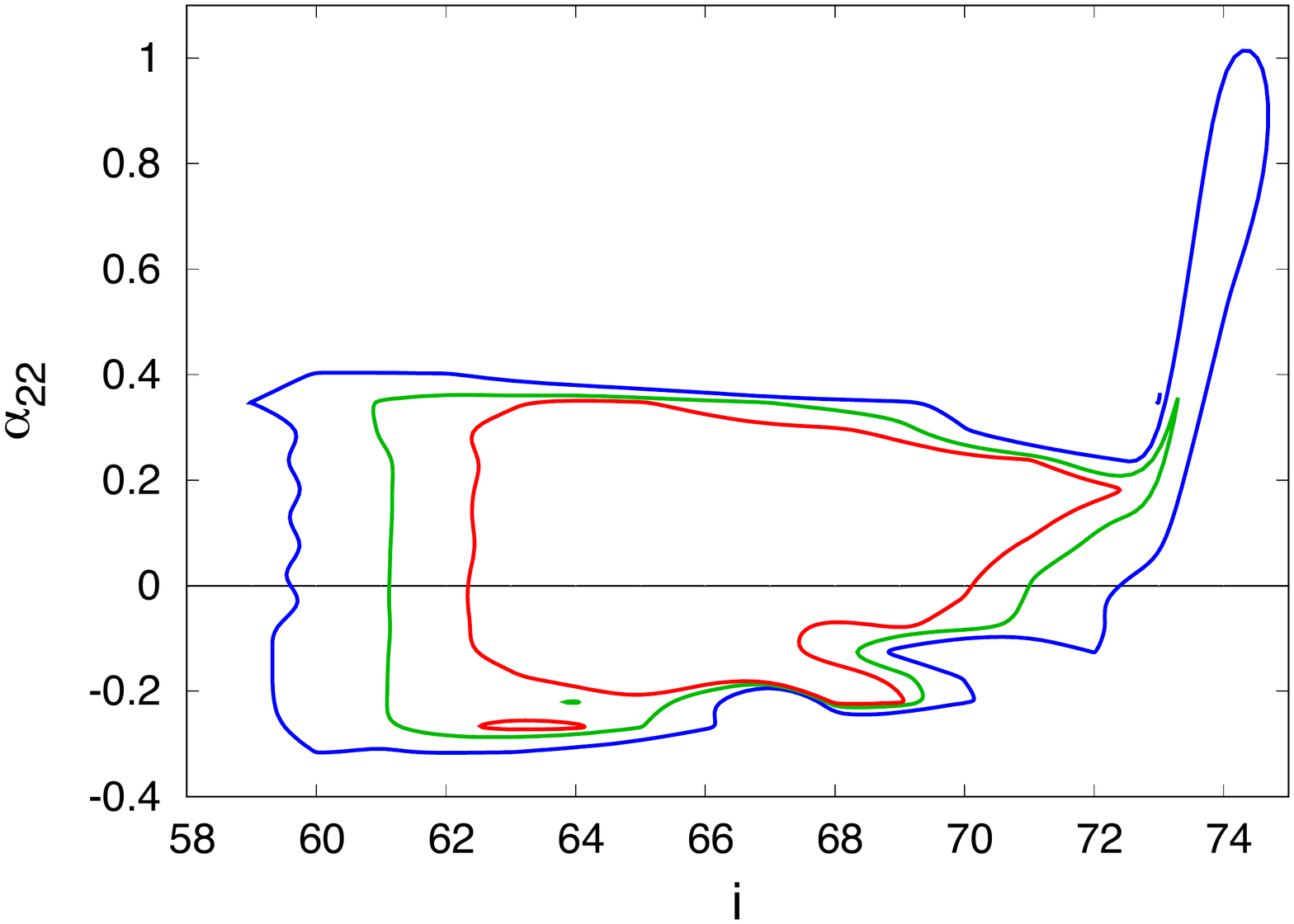}\label{incl}
        }
        
     \caption{Panel (a): Contour plot showing the degeneracy between $\alpha_{13}$ and $log~\xi$ assuming model D. Panel (b): Contour plot showing the degeneracy between $\alpha_{22}$ and $i$ assuming model D. The red, green and blue contour lines indicate 68\%, 90\% and 99\% confidence levels, respectively.} \label{pars}
\end{center}
\end{figure}

Apart from the fitting glitches (i.e., random fits getting stuck and the failure to yield bounds on the deformation parameters), some additional anomalous behavior were detected when fits would land in the \emph{far-Kerr} minima. In the $\alpha_{13}$ case, for several stepping resolutions, a random fit in the \emph{far-Kerr} minimum showed a drop in the disk temperature ($log~\xi < 1.0$). No other parameter was affected. To investigate further, we calculate the two-dimensional degeneracy between $\alpha_{13}$ and $log~\xi$. Fig.~\ref{logxi} shows this contour. We obtain closed contours with $log~\xi > 1.0$ with no hint of extension to lower values, suggesting that the \emph{far-Kerr} minimum with $log~\xi < 1.0$ are outliers. The uncertainty in $\alpha_{13}$ from this plot is much more consistent with the \emph{near-Kerr} contour rather than the \emph{far-Kerr} contour. In the case of $\alpha_{22}$, random \emph{far-Kerr} fits showed a drop in the inclination values. To investigate further, we compute the two-dimensional $\alpha_{22}-i$ degeneracy. This result is shown in Fig.~\ref{incl}. Here again we obtain closed contours. The uncertainty in inclination from this plot matches well with that obtained from Fig.~\ref{i-q} and the error in $\alpha_{22}$ is consistent with the \emph{near-Kerr} contour rather than the \emph{far-Kerr} contour of Fig.~\ref{A22Z}. This again suggests that the \emph{far-Kerr} contours cover a disconnected parameter space and are spurious.
\begin{figure}[!t]
\centering
\includegraphics[width=0.475\textwidth, trim={4mm 4mm 4mm 2mm},clip]{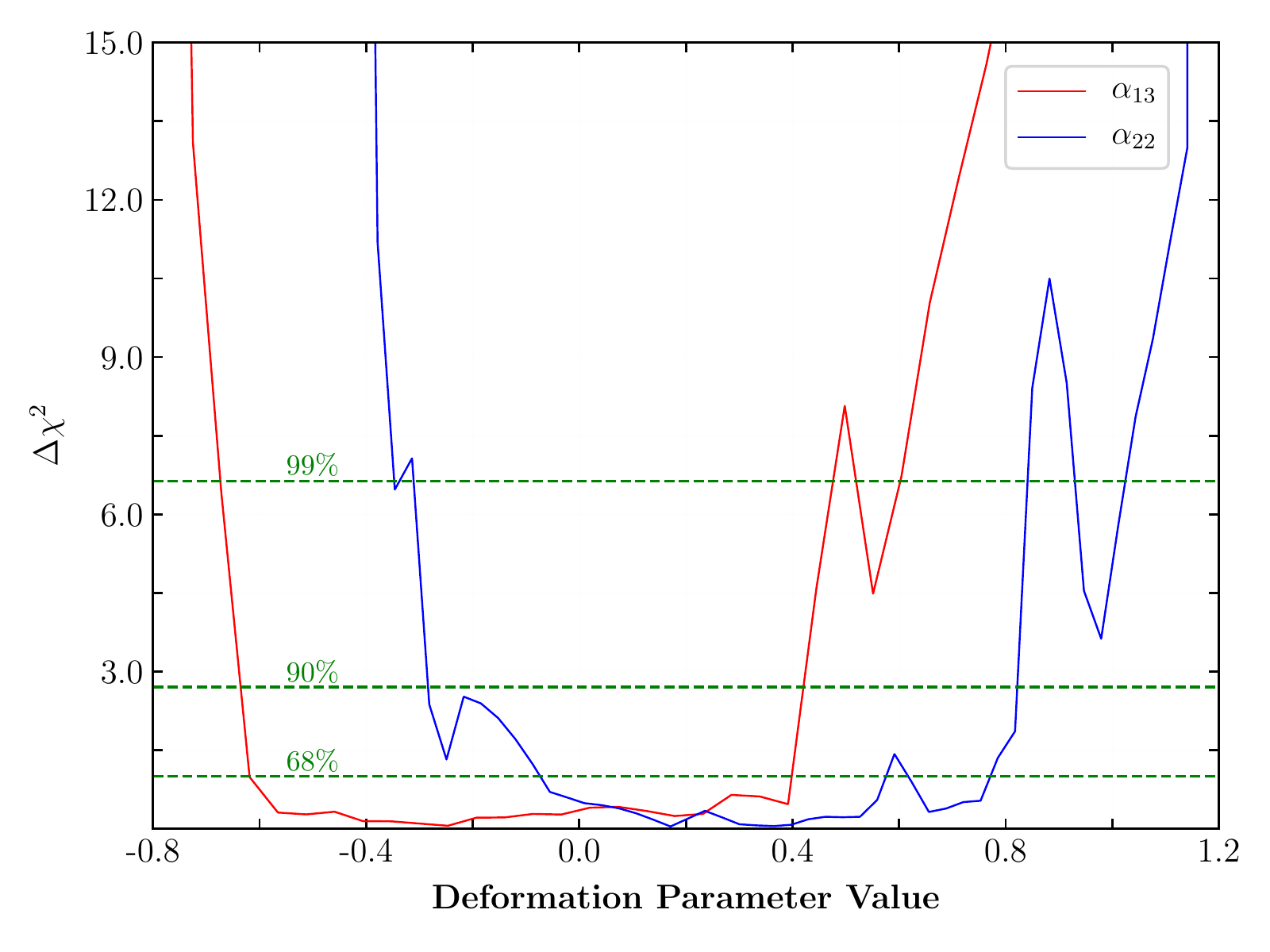}
\caption{Plot shows the one-dimensional contours of the deformation parameters $\alpha_{13}$ and $\alpha_{22}$ against the delta-fit statistic ($\Delta\rchi^2$) with indications for uncertainties at 1 parameter of interest under 68\%, 90\% and 99\% confidence intervals.}\label{1Dcont}
\end{figure}

We performed an additional test to check for the robustness of the uncertainty in deformation parameters. Instead of a two-dimensional \textsc{steppar} which shows the degeneracy of a deformation parameter with another model parameter, we perform a one-dimensional \textsc{steppar} with just the deformation parameter. This result is shown in Fig.~\ref{1Dcont}. Here we very clearly obtain a contour around the Kerr solution (corresponding to zero deformation parameter) and the range of deformation parameter is consistent with that obtained from the \emph{near-Kerr} contours of Fig.~\ref{ALPHA}.

The solutions were further subjected to a final test, by limiting the grids of physical parameters to sensible ranges (Tab.~\ref{ranges}) and checking the results for any difference. We discovered no startling change. Neither do the \emph{far-Kerr} solutions disappear from Fig.~\ref{ALPHA} nor do they appear in Fig.~\ref{pars}.\footnote{However, $log~\xi$ and $i$ displayed lower uncertainties on the \emph{near-Kerr} deformation parameter bounds, further defying the \emph{far-Kerr} solutions.} At this point we caution that we do not claim if the \emph{far-Kerr} solutions are physical or not, but simply describe what this low-reflection dataset shows us.

\begingroup
\begin{table}[!h]
\def\arraystretch{1.1}
\setlength{\tabcolsep}{1mm}
     \begin{ruledtabular}
     \caption{Examples of the limited domains of model parameters for the final cross-examination of the \emph{far-Kerr} solutions. All combinations of these sampled ranges were inspected with. Parameter units can be referred to from either Tab.~\ref{T1} or Tab.~\ref{T2}.}
     
     \begin{tabular}{ccc} \label{ranges}
     
     Component & Parameter &  Range(s) \\ \hline \hline 
      \multirow{6}*{\textsc{zpcfabs}} & $n_{\textrm{H}}$ & 1--1000 \\
      												 & & 1--100 \\
      												 & & 2--10 \\ \\
      												 
 													& $CvrFract$ & 0.0--0.8 \\ 
 													& & 0.1--0.5 \\ \hline 
 													
 	\multirow{2}*{\textsc{zpowerlaw}} & \multirow{2}*{$PhoIndex$} & 2.0--2.8 \\ 
 														 & &  2.3--2.6  \\ \hline
											
     \multirow{17}*{\textsc{relxill\_nk}} & $q$\footnote{Although \cite{W13} proposed limiting $q \geqslant 3$, we tested with flatter emissivities included and saw that they lead to contours similar to Fig.~\ref{zxipcf}.} & 4--10  \\
     													& &  5--10 \\
     													& &  6--10 \\
     													& &  1--10 \\ \\
     
     														& $a_*$ & 0.7--0.998 \\ \\
     														
     														& $i$ & 30--75 \\
     														& &  55--75 \\ \\
     														
 															& $log~\xi$ & 0.0--3.0 \\
 															& &  0.0--2.0 \\
 															& &  1.0--2.0 \\
 															& &  1.0--4.7 \\ \\
 															
 															& $A_{\textrm{Fe}}$ & 0.5--5.0 \\
 															& &  0.5--2.0 \\
 															
       \end{tabular}
       \end{ruledtabular}
\end{table}
\endgroup

\section{Ionized vs unionized partial covering} \label{app:absorb}

\begin{figure}[!t]
\begin{center}
        \subfigure[]{

             \includegraphics[width=0.45\textwidth, trim={1.3cm 0 4.3cm 2.7cm},clip]{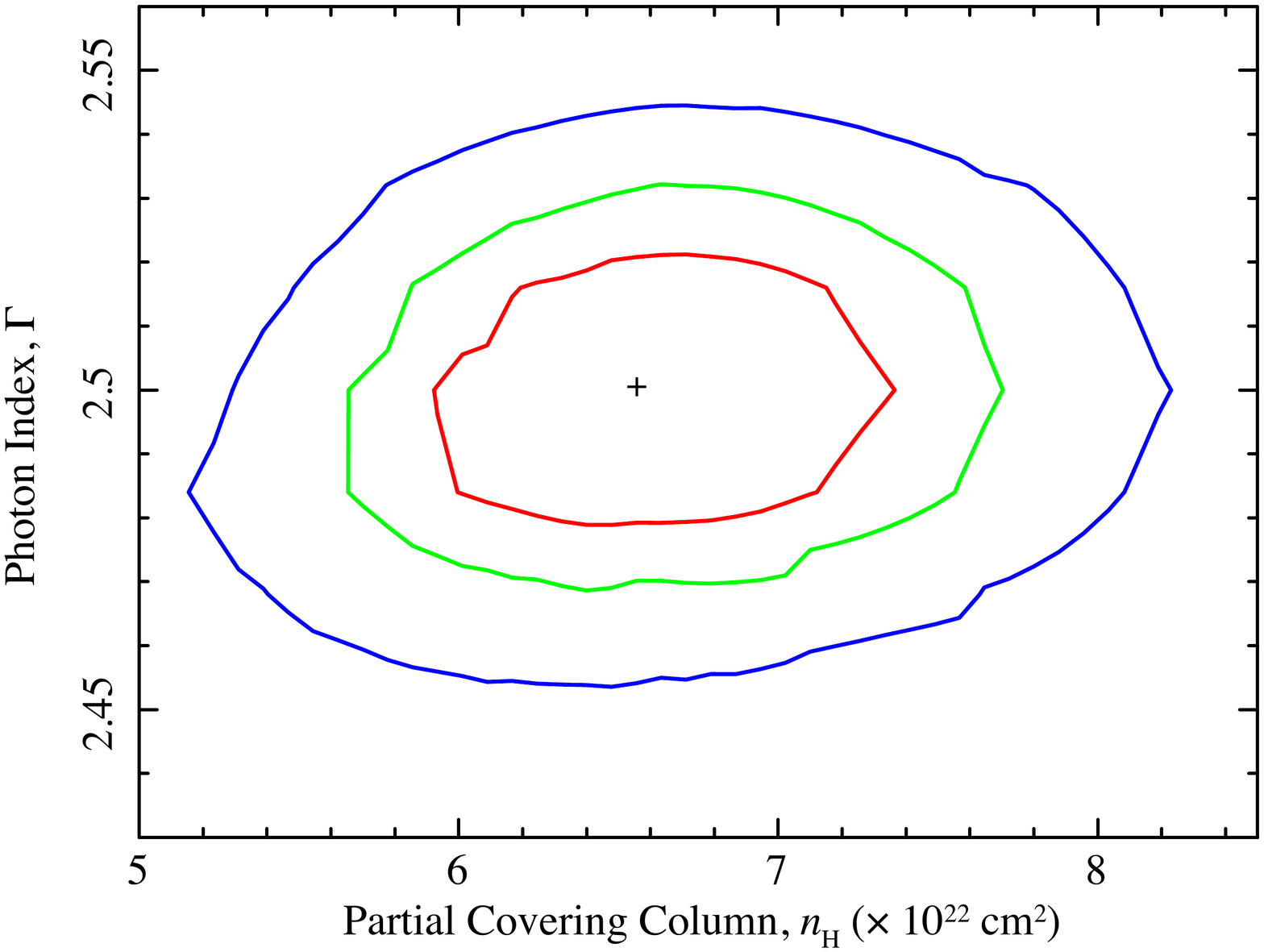}\label{C1}
        }
        \quad
        \subfigure[]{
        
           \includegraphics[width=0.45\textwidth, trim={1.3cm 0 4.3cm 2.7cm},clip]{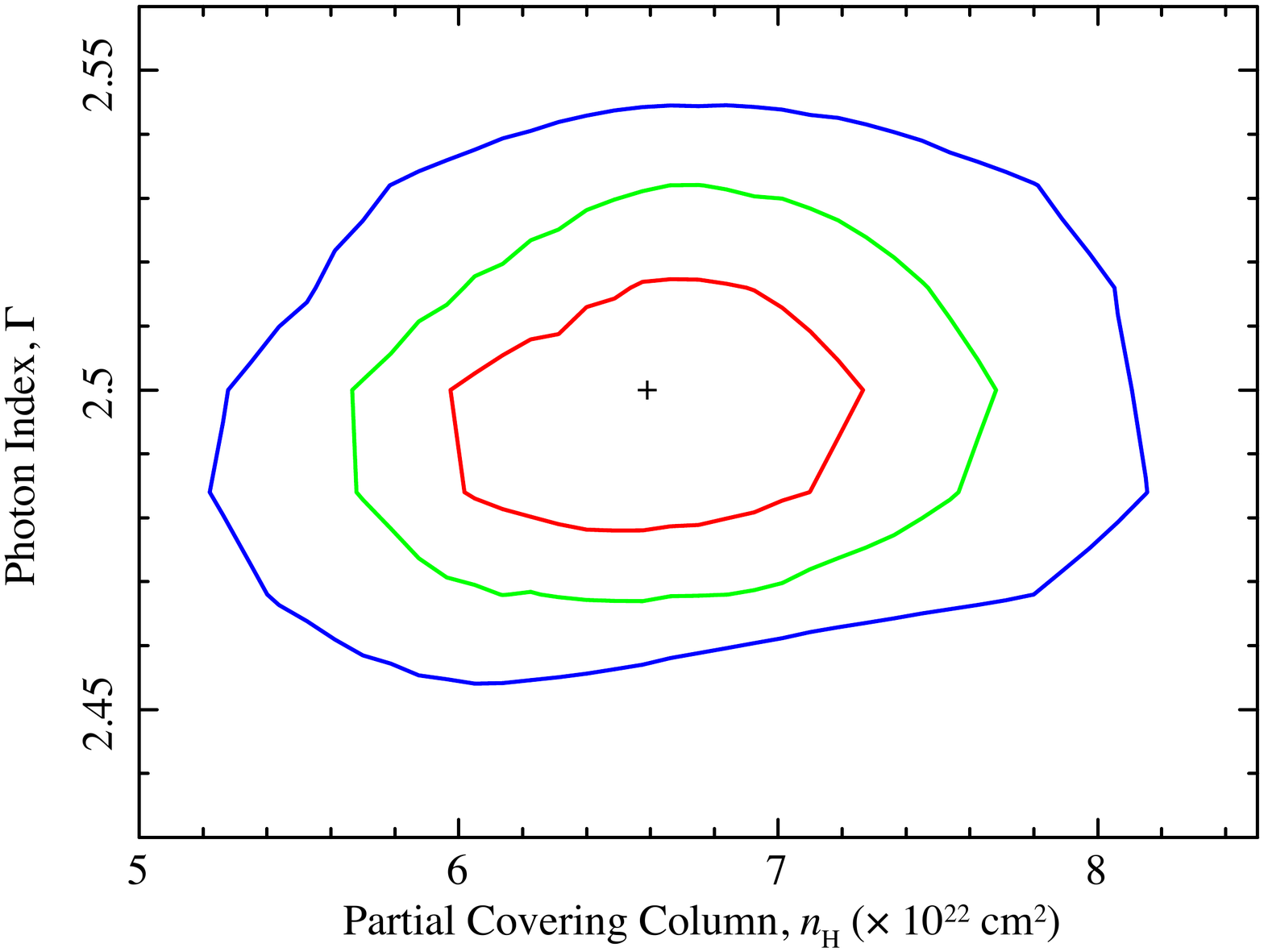}\label{C2}
        }
                
     \caption{Two-dimensional contour plots showing the degeneracy between the spectral photon index parameter $\Gamma$ versus the partial covering hydrogen column $n_{\textrm{H}}$ in Model D, for $\alpha_{13} \neq 0$ (left) and $\alpha_{22} \neq 0$ (right). Plot was rebinned on \textsc{xspec} for clarity.} \label{D2}
\end{center}
\end{figure}

\begin{figure}[!h]
\centering
\includegraphics[width=0.475\textwidth, trim={3.2cm 3cm 2.2cm 2.2cm},clip]{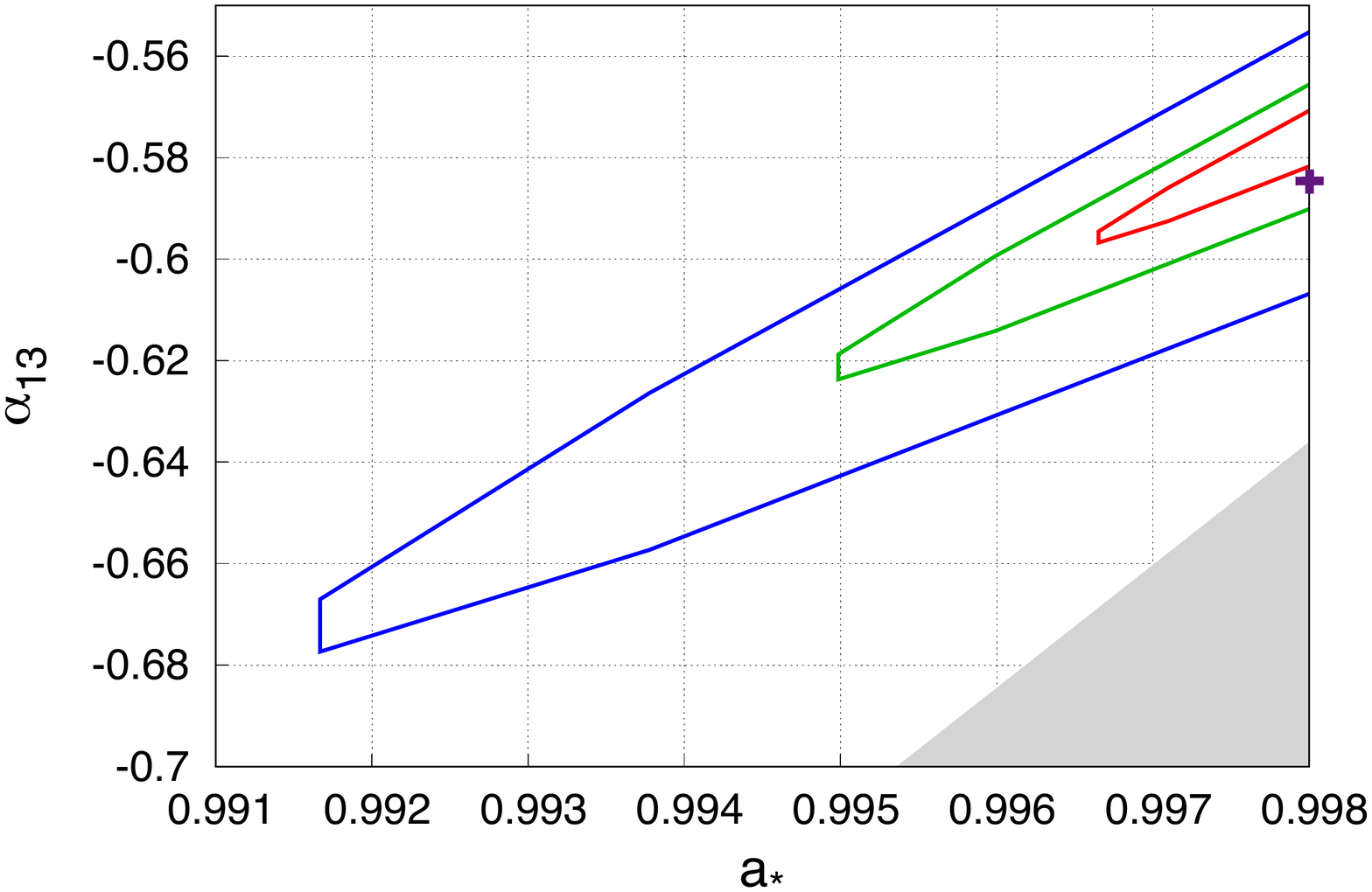}
\caption{Similar to Fig.~\ref{A13Z}, but using \textsc{zxipcf} instead of \textsc{zpcfabs}. The purple ``\textbf{+}'' marks the best-fit. Note that the Kerr solution ($\alpha_{13}=0$) is not recovered.} \label{zxipcf}
\end{figure}

When applying a partial covering absorber, it was not clear at the onset whether the absorber is ionized or unionized. We chose to model the ionized absorber with \textsc{zxipcf} and warm absorbers built on \textsc{xstar} \citep{XSTAR}. The unionized one was modeled with \textsc{zpcfabs}. With \textsc{zxipcf} we did manage to get a better fit over the use of \textsc{zpcfabs}, where the latter yielded the next-to-best statistical convergence among all model combinations tested with. However, no constraint was obtained for the hydrogen (\textrm{H I}) column $n_{\textrm{H}}$ of the ionized covering against the X-ray continuum index $\Gamma$. Among all absorbers, only the \textsc{zpcfabs} model provided good constraints on $n_{\textrm{H}}$ (see Fig.~\ref{D2}), which indicates that the model is statistically restrained. Moreover, it is \textsc{zpcfabs} that produced decent constraints on both deformation parameters (Fig.~\ref{ALPHA}) with comparatively 1 less d.o.f. than \textsc{zxipcf}. Fig.~\ref{zxipcf} shows the constraint obtained with Model D if \textsc{zxipcf} was used instead. The model completely excludes the Kerr solution.
The fit statistic is exceptionally good in this case ($\Delta\rchi^2$/d.o.f. = 986.78/958 $\sim$~1.03). Naively, one may get mislead by such statistics. But we have seen in other instances, e.g.,~\cite{Xu2018}, that some model degeneracies result in exceptional fit statistics at highly non-Kerr solutions (i.e., excluding the Kerr solution at very large $\sigma$), but are usually a result of incorrect modeling. Moreover, we do not obtain any closed contour for the $\alpha_{22}\neq0$ case with \textsc{zxipcf}. We, thus, conclude in favor of \textsc{zpcfabs}.

\bibliography{Mrk335SuzakuNK_ref}

\end{document}